\documentclass[]{aa}  

\usepackage{cancel}
\usepackage{subfig}
\usepackage{graphicx}
\usepackage{amssymb}
\usepackage{txfonts}
\begin{document} 

\titlerunning{Diffuse interstellar gas in GALEX AIS}
\authorrunning{Armengot \& G\'omez de Castro}

   \title{Signatures of  diffuse interstellar gas in the  Galaxy Evolution Explorer all-sky survey\footnote{Table 1 is only available in electronic form
at the CDS via anonymous ftp to \url{cdsarc.u-strasbg.fr} (130.79.128.5)
or at \url{http://cdsweb.u-strasbg.fr/cgi-bin/qcat?J/A+A/}}}

   \subtitle{}

   \author{Marcelo Armengot\inst{1} \and Ana I. G\'omez de Castro\inst{1}}

   \institute{AEGORA Research Group, Facultad de CC Matem\'aticas, Universidad Complutense de Madrid, 28040 Madrid, Spain\\
\email{aig@ucm.es}}

   \date{Received \today}

 
  \abstract
      {     
        The all-sky  survey  run by the Galaxy Evolution Explorer (GALEX AIS)  mapped  about 85\% of the Galaxy  at ultraviolet (UV) wavelengths  and detected  the diffuse UV background produced by the scattering of the radiation from OBA stars by interstellar dust grains. Against this background, diffuse weak structures were detected as well as the UV counterparts to nebulae and molecular clouds.       }
      {
        To make full profit of the survey, unsupervised and semi-supervised procedures
       need to  be implemented. The main objective of this work is to implement and  analyze the results of  the method developed by us for the blind detection of ISM features  in the GALEX  AIS .       }      
      {
        Most ISM features are detected at very low signal levels (dark filaments, globules)  against the already faint UV background. We have defined an index, {\it the UV background fluctuations index} (or UBF index), to identify areas of the sky where these fluctuations are detected. The algorithm is applied to the images obtained in the  far-UV (1344 -- 1786 $ \AA $)  band since this is less polluted by stellar sources, facilitating the automated detection. 
      }
      {    
     The UBF index is shown to be sensitive to the main star forming regions within the Gould's Belt, and to some prominent loops
     like Loop I or the Eridanus and Monogem areas.  The catalog with the UBF index values is made available online to the community.  
                   }


   \maketitle
%

\section{Introduction}

In 2003, the GALaxy Evolution eXplorer ({\em GALEX}) was launched  into low Earth orbit. The mission lasted ten years
and was equipped with wide-field  cameras that provided simultaneous  far-ultraviolet (FUV) 
(1344 -- 1786 $ \AA $)   and near-ultraviolet (NUV) (1771 -- 2831 $ \AA $) imaging using photon-counting MCP-type detectors
(Martin et al. 2005).  GALEX ran the first all-sky survey at UV wavelengths providing highly valuable data for interstellar 
medium (ISM) studies.  

The UV radiation produced by the massive OBA stars in the Galaxy is scattered and absorbed by the dust  in the ISM
providing important clues to the  albedo and composition of the grains (Saslaw \& Gastaud 1969, Draine, 2003), especially at 
the low   end of the size distribution. An additional contribution to the UV ISM radiation comes from the molecular 
hydrogen in the envelopes of the 
molecular clouds that efficiently absorbs  the Lyman-$\alpha$ photons from the UV background and reprocess them into H$_2$ 
fluorescent emission in the electronic Lyman band; these transitions are significantly stronger than the infrared 
(magnetic quadrupole) transitions often used in ISM studies. Moreover,  the envelopes of ionized nebulae are detected 
in the GALEX bands;  the resonance transitions from neutral (H~I, C~I, O~I) singly ionized (Mg~II, C~II, O~II, 
Fe~II, Al~II) or multiply ionized gas (C~IV, Si~IV, N~V) are observed within the range covered by GALEX.

Some examples of GALEX images of interstellar objects are shown in Figures 1-3.
The Rosette nebula is an excellent example of a reflection nebula. Rosette is illuminated by  a cluster of B8-9 stars, and
dense (dark) filaments are observed against the bright background produced by the  radiation  scattered by the  icy dust grains
in the  nebula (see Figure~1). 
 
        \begin{figure}[h]
             \includegraphics[width=9cm]{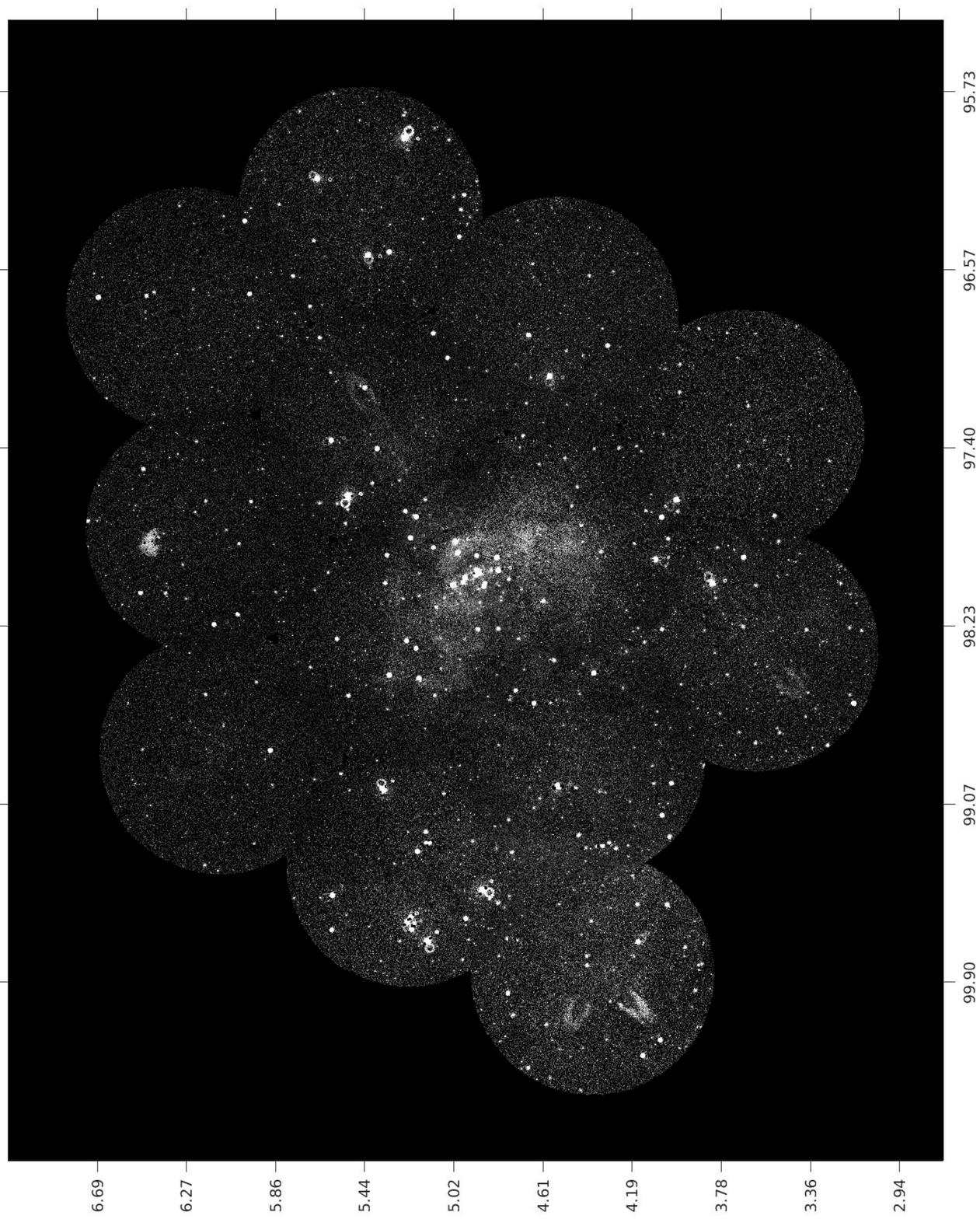}\\
             \includegraphics[width=9cm]{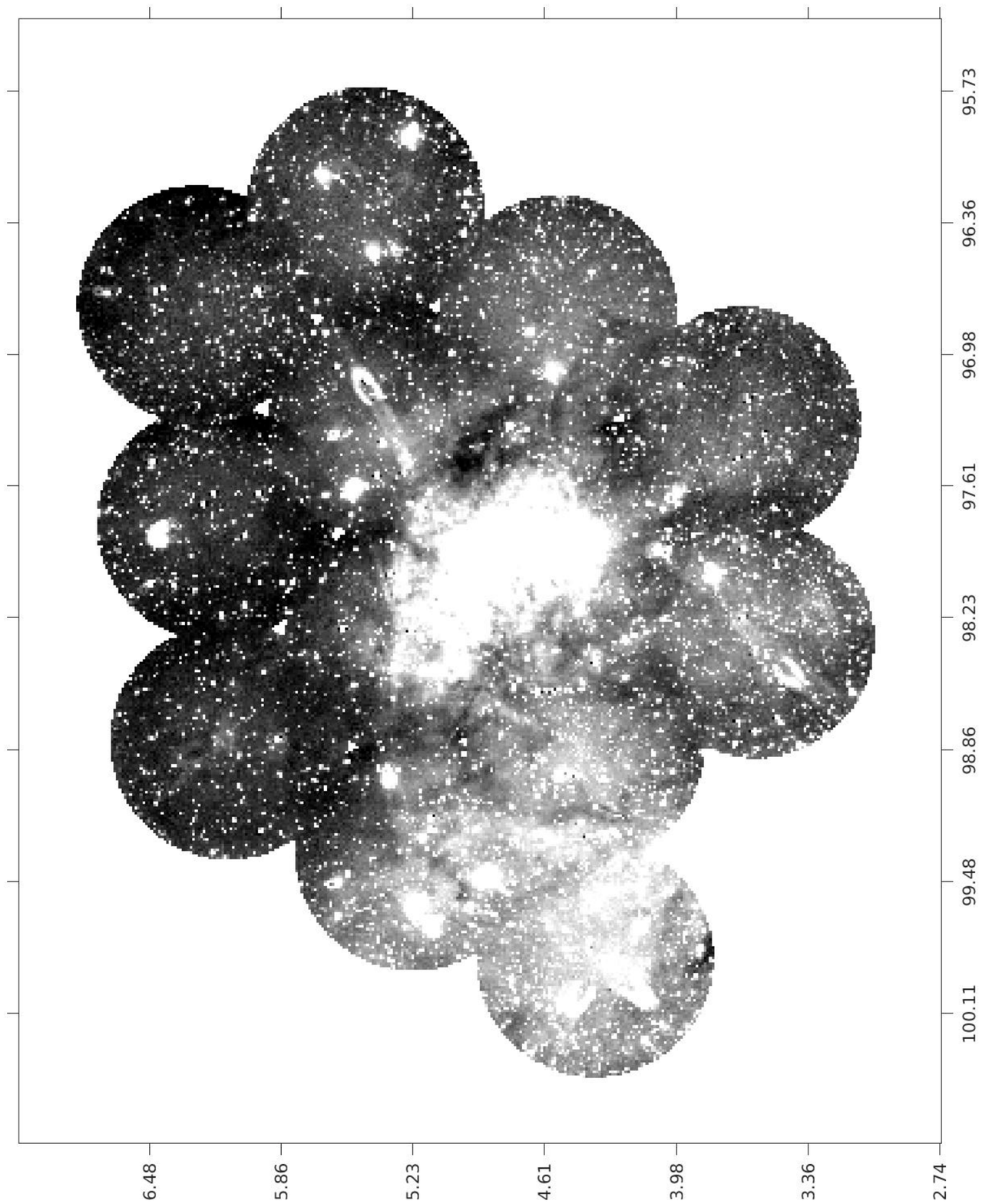}
          \caption{Rosette nebula as seen by GALEX in the NUV band (Rosette  was not mapped in the
            FUV band). The image is a composite of 13 NUV images assembled using the MOSAIX tool (Armengot et al. 2014). {\bf Top panel:} High contrast image. {\bf Bottom panel:}
            Same image in logarithmic scale to enhance the  structures at low signal level. Coordinates are ICRS: right ascension increases from right to left (in degrees) and
            declination from bottom to top.}
          \label{roseta}
        \end{figure}

        GALEX sensitivity to faint absorption nebula is shown in Figure~2 for Barnard 29. The exposure time is about 200 s 
        in the FUV band, significantly less than required to obtain an image of the dark clump with similar S/N in the optical range.

        \begin{figure}[h]
          \centering
          \includegraphics[width=6cm]{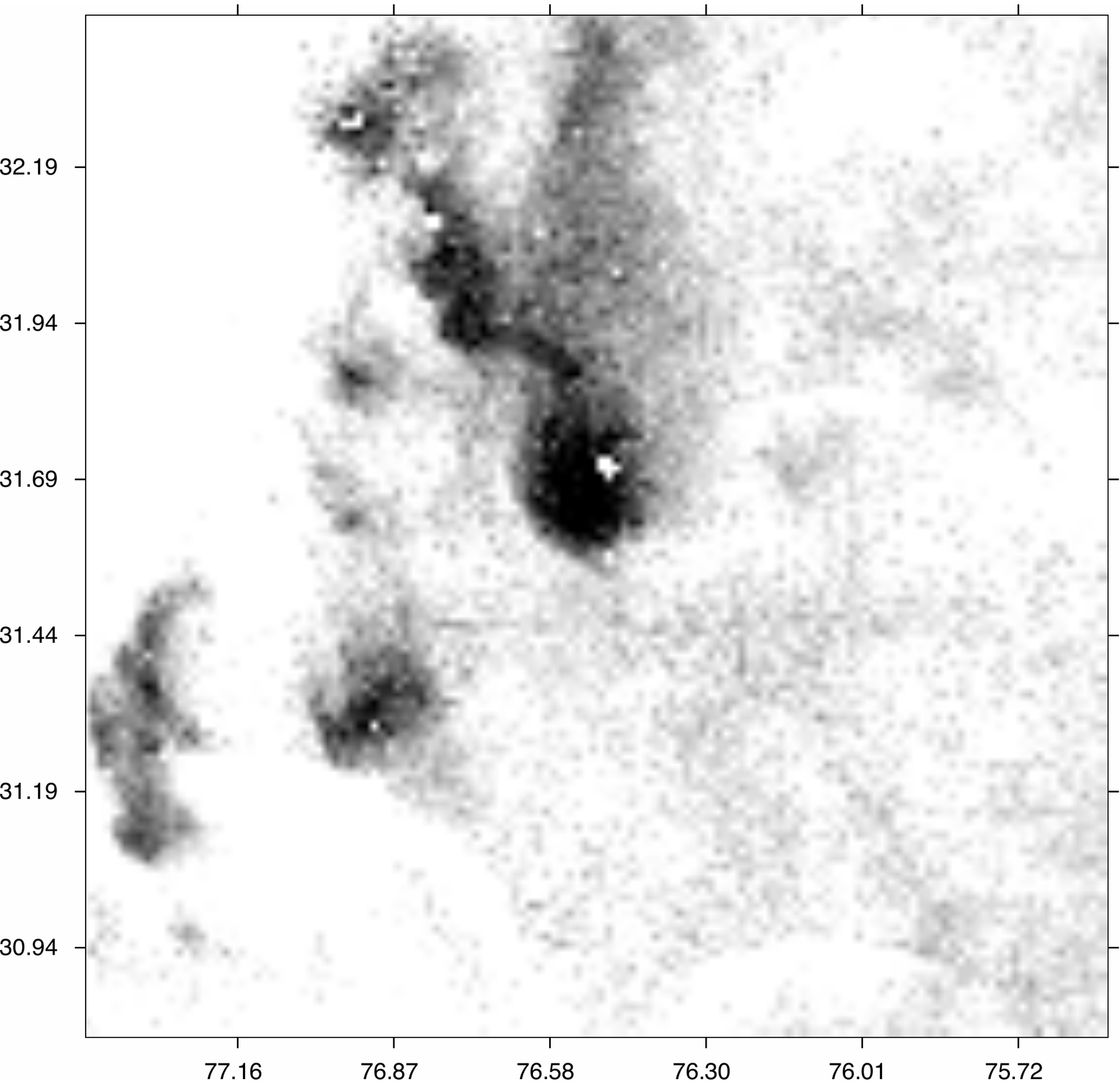}
          \caption{Barnard 29  imaged  in the FUV band by GALEX AIS.
            The original GALEX images have been mosaiced together (coordinates as in Figure 1).}
          \label{3clouds}
        \end{figure}

        The Vela loop is shown in Figure~3. The loop is prominent both in the NUV  (Fe~II, C~II)
        and in the FUV (C~IV, Si~IV, He~II) bands and the braiding of the filaments is   neatly observed
       even in the $\sim 90$~s exposure time images of the GALEX All-Sky Imaging Survey (AIS).
        
          \begin{figure}[h]
            \centering
            \includegraphics[width=8cm]{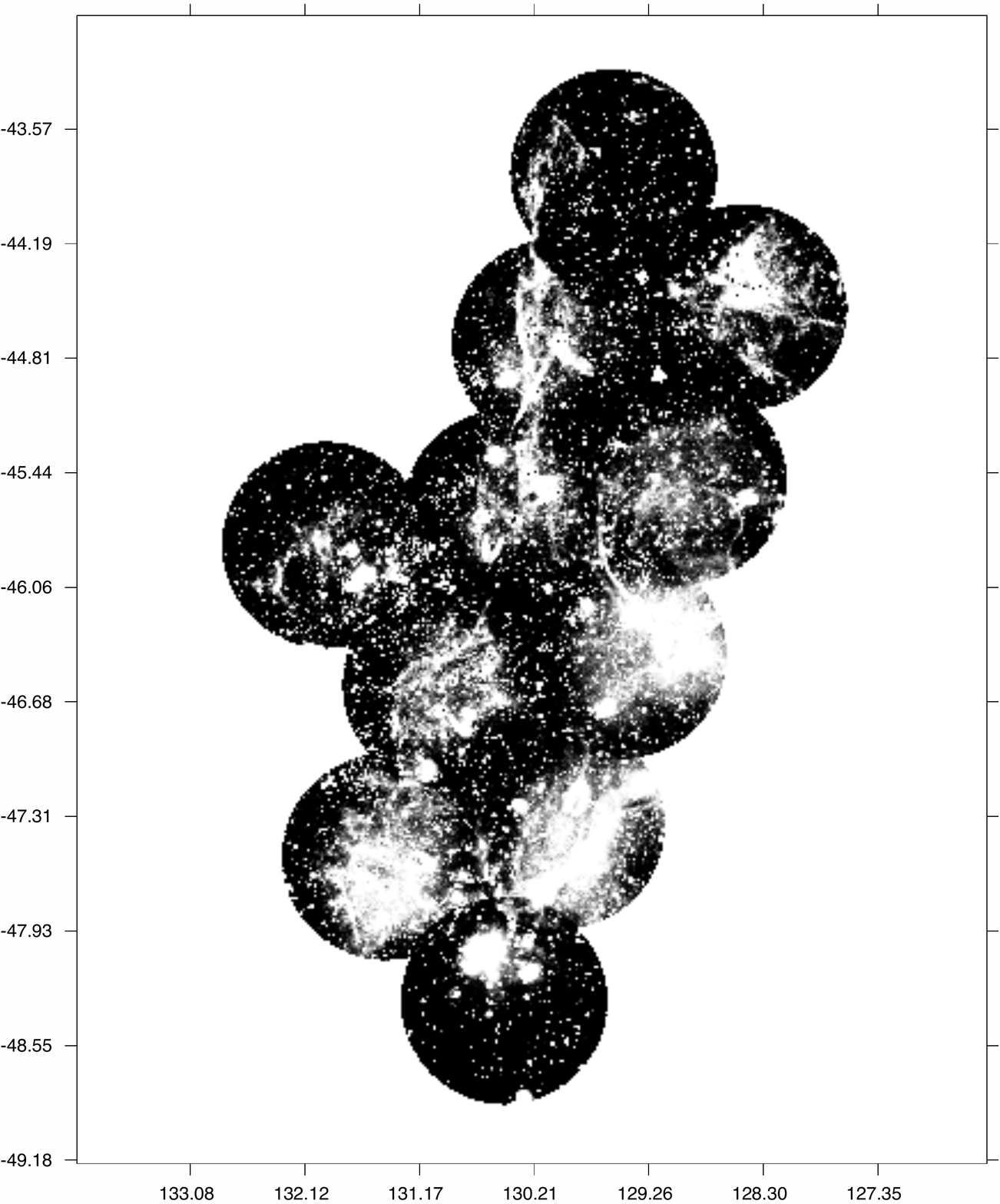}
            \caption{Vela filaments observed by GALEX in the NUV band. 
The image is a composite  of 11 NUV images (coordinates as in Figure 1).}            
            \label{vela}
          \end{figure}
          
          All these features were observed at intensity levels  similar to the UV background level. 
          The dark globules in Figure~2 would have passed unnoticed to a simple inspection of the GALEX tiles.

        During its lifetime, the GALEX AIS  produced  28,707 images (or tiles) in the FUV band and 34,285 images in the NUV band (Bianchi et al. 2014), with a total sky coverage of 26,000 square degrees.  The wealth of information contained in GALEX AIS images is huge and requires  unsupervised procedures to search for weak interstellar structures.
This work deals with this challenge; we describe, implement, and analyze the results of  the method we developed   for the blind detection of nebular structures, condensations, and filaments in the GALEX survey. In Section 2 we describe the method and derive the index to search for  diffuse features in the GALEX AIS. In Section 3 the results are presented and compared with other galactic surveys. The article concludes with a short summary.
        
\section{Description of the method} 
\label{BSVR}

\begin{figure*}[h!]
\begin{center}
\includegraphics[width=12cm,angle=270]{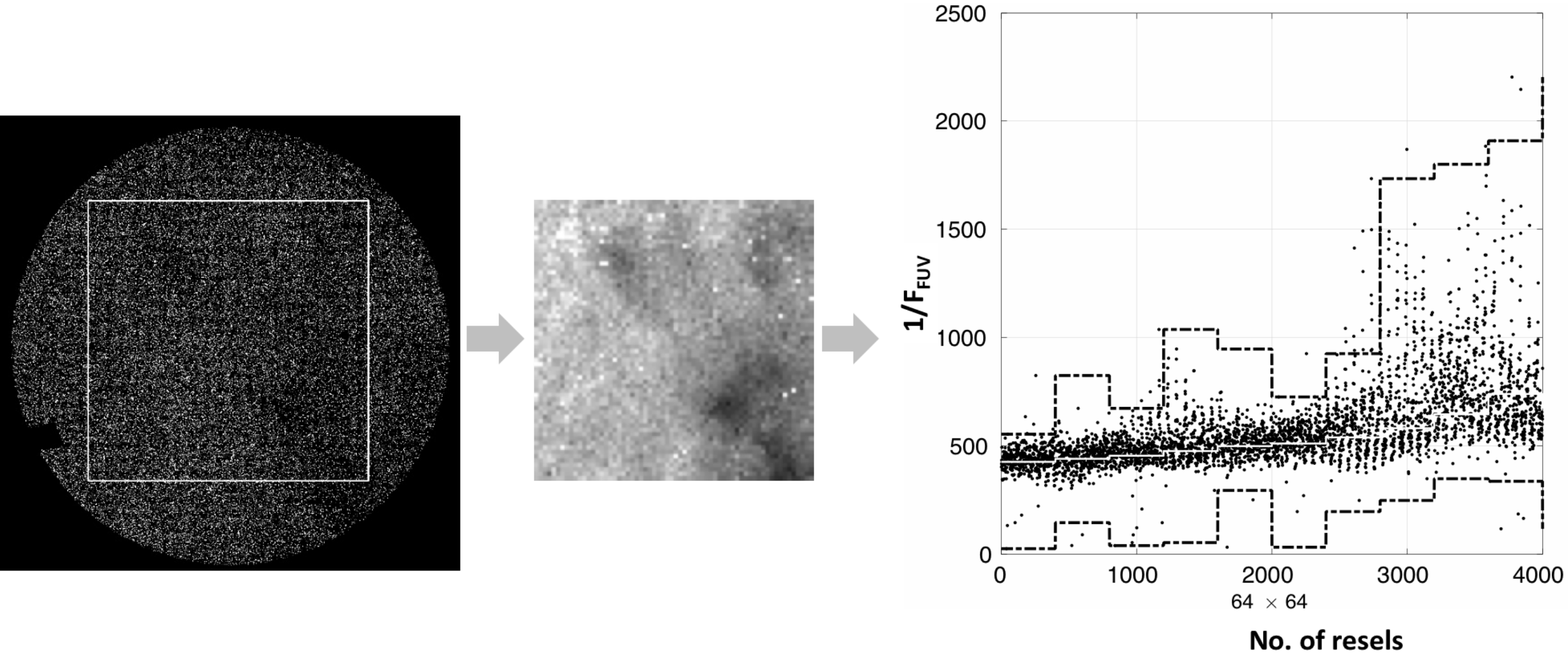}\caption{Procedure to compute the BSVR. A sample GALEX image is displayed in the leftmost panel, as well as the central square used to compute the BSVR. This central square rebined into an image of  $64 \times 64$  resels which is displayed in the central panel. From the inverse of this image, a vector,  $V$, is constructed by re-arranging its elements (see text); they are plot in the right panel. The envelopes in this panel are constructed as follows: every 400 elements, the mean (white line), the maximum,  and the minimum (bold dashed lines) are computed. The BSVR is determined from these values (see text for further details).}
\end{center}

\label{hiperdata}
\end{figure*} 
        For each field imaged by GALEX, the mission provides detailed information including  raw data,  calibration files, and the end data products (see Morrisey 2007), all accessible through the Mikulski Archive of Space Telescopes\footnote{archive.stsci.edu} (MAST). 
 Among the whole set of data products provided for each field or tile,  there are three relevant to this work: 

          \begin{itemize}
          \item[(1)] The flux calibrated intensity map (extension  {\bf \em -int } in the archive);
          \item[(2)] The sky background in the area as modeled by the mission and implemented in the data  pipeline (extension {\bf \em -skybg} in the archive); 
          \item[(3)] The final background subtracted image (extension {\bf \em -intbgsub} in the archive).
          \end{itemize}

          Upon the  completion of the mission, GALEX AIS data products were used to study the Galactic UV background at the Sun's location in the Galaxy. Murthy et al.  (2010) built a map of the local  UV background  at NUV and FUV wavelengths using the sky background data derived by the mission (the {\em \bf -skybg} files). An all-sky map of the UV background was provided after normalization by the exposure time and mosaicking the full set of  {\em \bf -skybg} images into an Aitoff projection.  Later work by Handem et al. (2013) applied a more sophisticated treatment that involved the removal of  the point-like UV sources from the intensity images ({\em \bf -int }) by implementing {\em mask} files including the effect of dust grains scattering of the UV radiation from point-like sources. For our work, we used the end data products from the mission, the calibrated images  ({\em \bf -intbgsub} images) distributed through the MAST.

                The first challenge for the blind detection of ISM features is their low signal;  ISM filaments are  observed as small  fluctuations over the background that it is already very weak.  For instance, the FUV background is $< 1 \times 10^{-2}$ ~counts~s $^{-1}$~pixel$^{-1}$ in the Taurus molecular complex (${\rm b}_{\rm gal} \simeq -20^o$) hence, less than one photon is detected per pixel in the $\sim 100$~s of exposure time, characteristic of GALEX AIS (Morrisey et al. 2007).   
To reach statistically significant levels, the images have been re-binned to resolution elements ({\it \emph{resels}}) of $30 \times 30$ pix$^2$ (or $45 \times 45$~arcsec$^2$); in this manner, the count rate is raised to tens of counts~s$^{-1}$ per resel even at high galactic latitudes. We 
note that this degraded angular resolution is still significantly better than the resolution of the IRAS survey at 100~microns (Beichman et al. 1998).  
This binning is similar to that used  by the GALEX mission to evaluate the UV background\footnote{\url{http://www.galex.caltech.edu/researcher/techdoc-ch3.html#4}}.

          After rebinning, the images were inverted \footnote{Each pixel is assigned a new value
          that is the inverse of the original; each pixel $(i,j)$ of the inverse image $M$ is assigned a value $m_{i,j} = 1/gr_{i,j}$, with 
          $gr_{i,j}$ the value of the pixel in the original GALEX rebinned image $GR$} to enhance the background. 
The statistical estimator of the background fluctuations is computed over these rebinned and inverted images $M$ 
with $64 \times 64$ elements $M(i,j)$ as follows: 

\begin{enumerate}

\item All elements of matrix $M$ are arranged into a vector $V$ of $64 \times 64 = 4096$ elements such that element $(i,j)$ in the matrix
corresponds to element $k$ in the vector with $k= (j-1)*64+i$;
\item Subsets $V_l$ with $l \in [1,10]$ are defined from $V$, so that $V_l = [ V((l-1)*400),V(l*400-1)]$, each containing 400 elements; 
\item For each subset $l$ the arithmetic mean, the maximum, and the minimum are computed: $Mean(l), Maximum(l)$, and $Minimum(l)$, respectively; an outlier detection algorithm was used; 
\item For each tile the background scale variability rating (BSVR)  is computed. The BSVR is defined as 
        \begin{equation}
          BSVR =  \frac {\sum _{l=1}^{l=10} (Maximum(l)-Mean(l))}{\sum _{l=1}^{l=10} (Mean(l)-Minimum(l))}     
        ,\end{equation}
        
          with $l$ running from 1 to 10, the number of subsets.
\end{enumerate}

The process is illustrated in Figure~4. The original GALEX image, as provided by MAST, is shown in the leftmost panel. Only the central square (1920$\times$1920 pix$^2$), 
marked in white in the image, is used for the calculation of the BSVR. This area is rebinned into resels of $30 \times 30$~pix$^2$, as shown in the central panel. Then the value of each pixel is plotted in 
the right panel; pixels are read from top to bottom and from left to right. Thus, the first 1,000 pixels correspond to the leftmost area
of the image where no significant fluctuations (filaments) are detected, while the last 1,000 pixels correspond to the rightmost
area where the dark filaments are concentrated. This results in an increasing dispersion from the top left corner (pixel number 1) to the bottom right corner (pixel 4,000). 

The BSVR is calculated over the background-subtracted images {\bf \em -intbgsub} files. Therefore, the BSVR is not be affected by any local or galactic
UV background; even cirrus or the dust scattered light from bright sources should be removed in the pipeline processing (see, e.g., figure 5 in \url{http://www.galex.caltech.edu/researcher/techdoc-ch4.html}). However, as shown in Figures 1-3, GALEX image processing does not remove reflection nebulae or dark clouds absorbing the UV background radiation. These are the targets to be detected using the BSVR.

We note that as $M$  is made of the inverse (rebinned) values, the contribution to the BSVR of the bright points or areas in the image  (mainly point-like sources) is negligible, provided that there are not many of them   and so do not affect significantly the mean. In this case, 
$Mean(k) >> Minimum (k) $, resulting in
\begin{equation}
{\rm BSVR} \rightarrow  \frac {\sum _{l=1}^{l=10} Maximum(l)}{\sum _{l=1}^{l=10} Mean(l)} -1
\end{equation}
The number of point-like sources in the FUV band is one-tenth of those detected in the NUV band.  Henceforth, the BSVR index is determined 
from  the FUV images in the survey to optimize its performance. 

The resulting BSVR all-sky map is plotted in Aitoff projection in Figure 5. Two main features are clearly discernible: [1] the BSVR depends on the galactic latitude and [2] there are some clear stripes (maximum circles) where the BSVR is anomalously low.
\begin{figure}
\centering 
\includegraphics[width=12cm,angle=270]{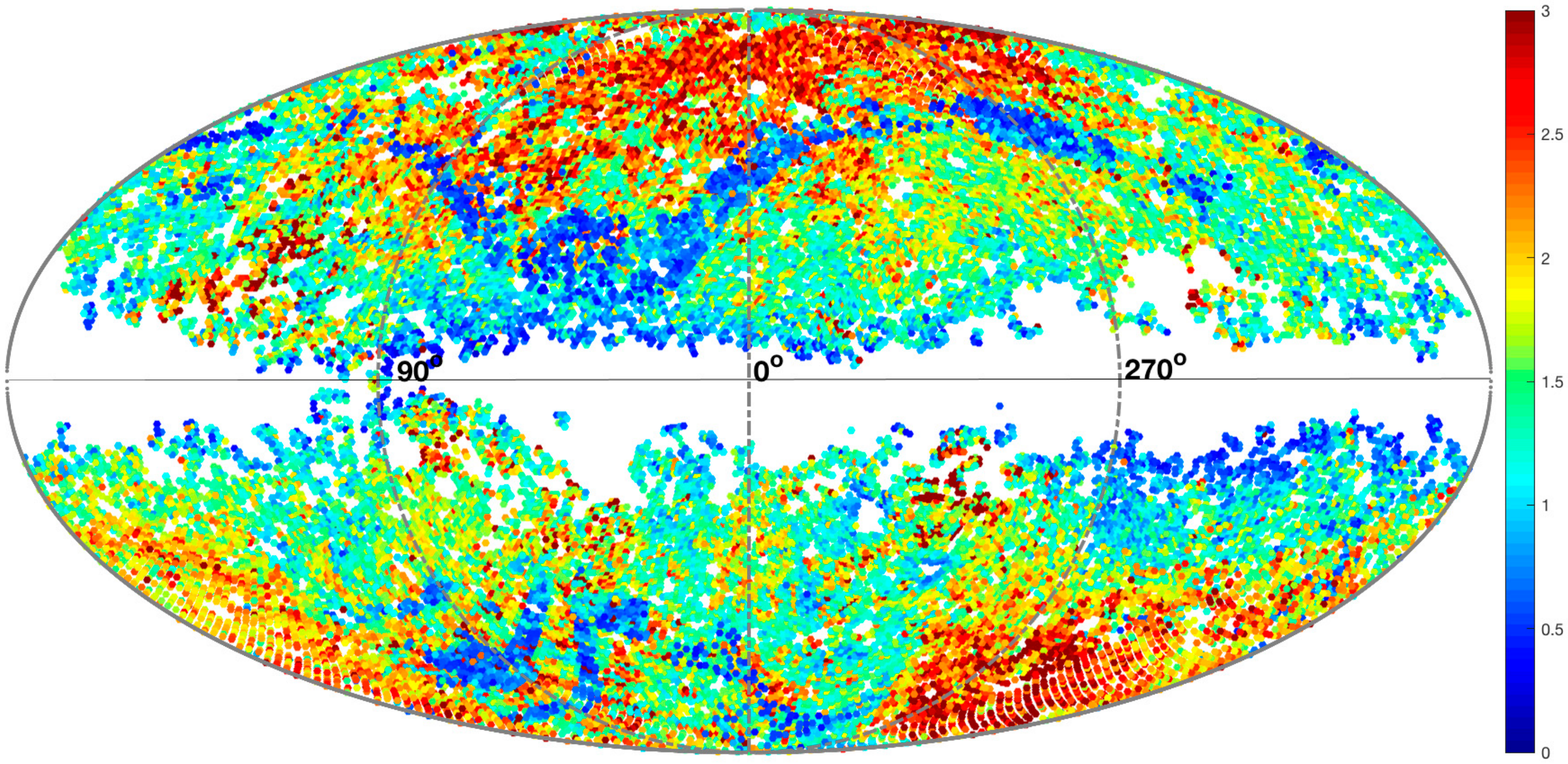}\\
\includegraphics[width=12cm,angle=270]{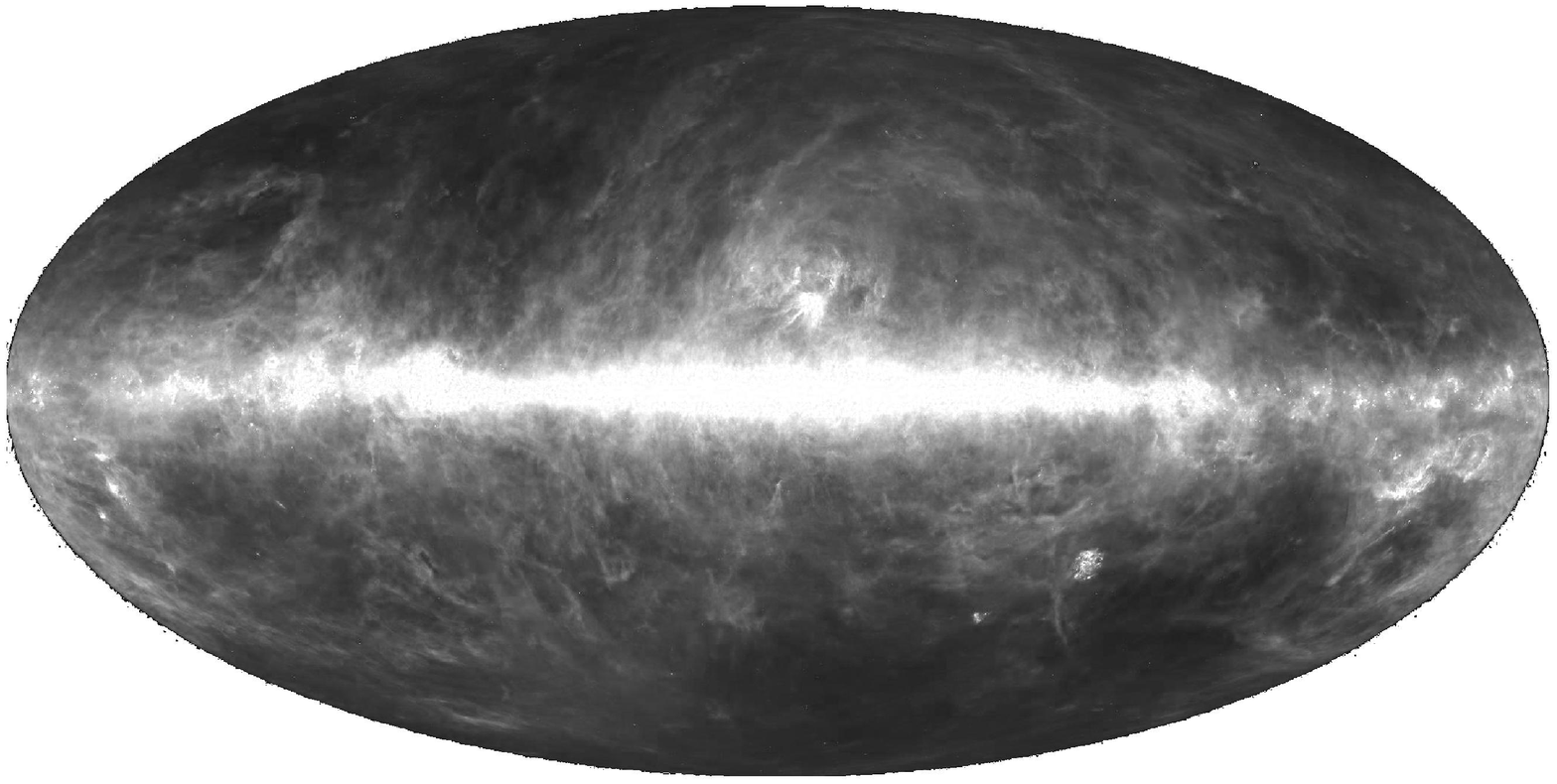}\\
\caption{{\bf Top:} Representation of  the BSVR for the 83.081 {\em GALEX} AIS images processed in this work; data are plotted
in Aitoff projection. {\bf Bottom:} IRAS 100-micron map of the sky displayed for reference.}
\label{themap}
\end{figure}

\subsection{Correction for the dependence on galactic latitude: the UV Background Fluctuations Index}

In Figure 6 the histogram of observed BSVR values per galactic latitude is represented. The butterfly  pattern
indicates that there are two neatly defined areas in the sky. Over most of the sky, the BSVR increases with galactic latitude;
however,  the area  with BSVR $< 0.6$ does not follow the same trend.

                     \begin{figure*}[h!]
             \centering
                       \includegraphics[width=12cm,angle=270]{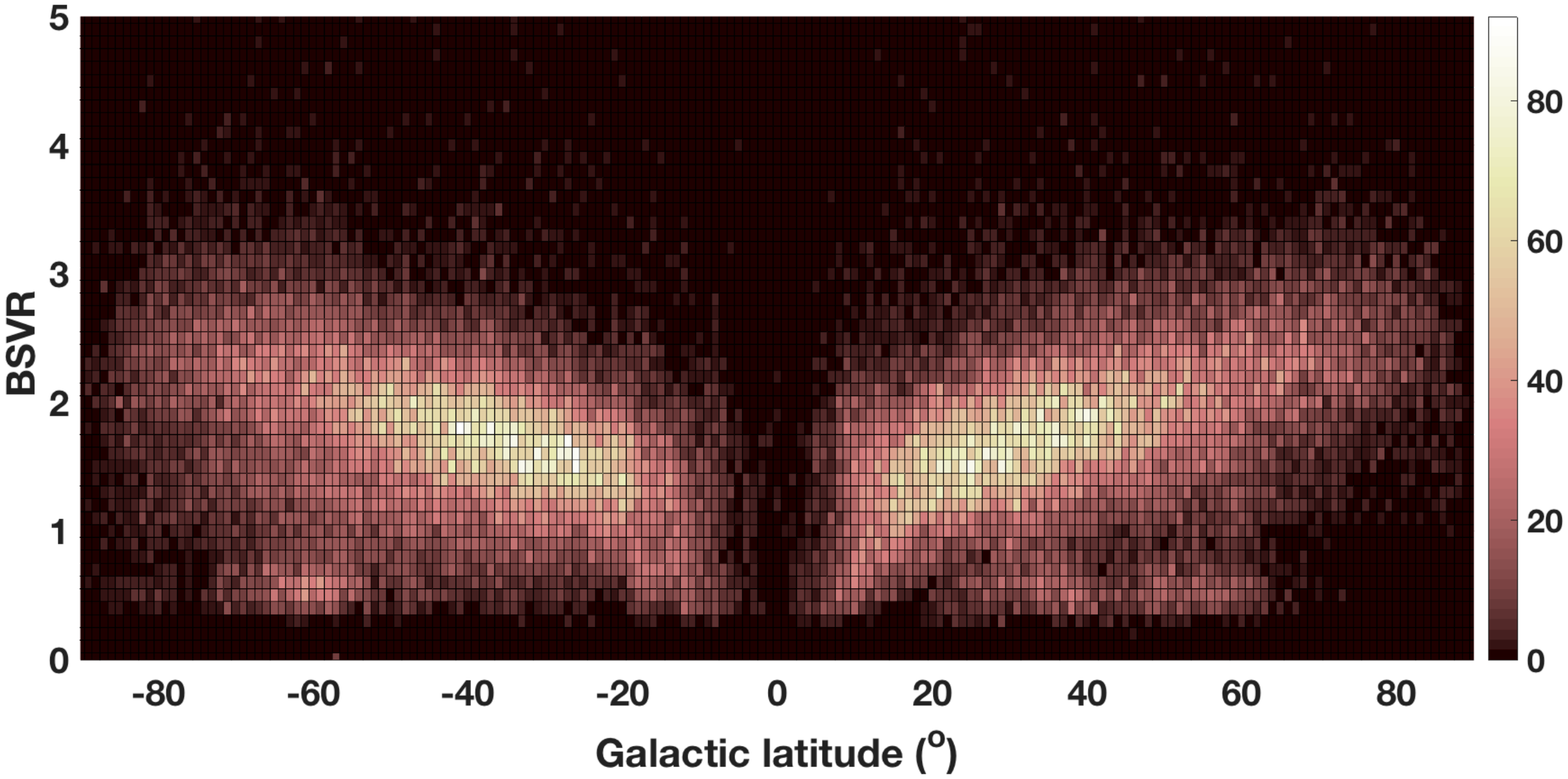}
                       \caption{ BSVR index plotted vs. galactic latitude for all tiles in the GALEX AIS survey.}
                       \label{UBFI_gallat}
                     \end{figure*}

The general dependence of the BSVR on galactic latitude is well fitted by a power law of $\sin |b|$,
\begin{equation}
BSVR =  2.070^{+0.005}_{-0.005} \times (\sin |b| )^{0.328\pm 0.004}
\label{fit}
,\end{equation}
\noindent
with RMSE=0.1185. The trend in Eq. 3 indicates that the BSVR is enhanced at high galactic latitudes
with respect to low galactic latitudes. The examination of the tiles at high galactic latitudes shows no enhancements 
in the number of nebulae with respect to the lower latitude tiles; in fact, the opposite trend is observed.
Moreover, as the BSVR is computed from the background subtracted files, it is not directly affected by the 
well-known dependency of the UV background on galactic latitude (Murthy et al. 2014). However, the very low signal 
level in the observations at high galactic latitudes makes the BSVR very sensitive to the low level 
statistical fluctuations introducing an indirect dependency in the background level and the galactic latitude. This
trend is not of astronomical origin, and here we  introduce the UV Background Fluctuations Index (UBFI) 
to correct for it. The UBFI is defined as
 \begin{equation}
{\rm UBFI }=  \frac{BSVR} {2.070 \times (\sin |b| )^{0.328}}
\label{fit}
\end{equation}

           \begin{figure}[h]
             \centering
                       \includegraphics[width=12cm,angle=270]{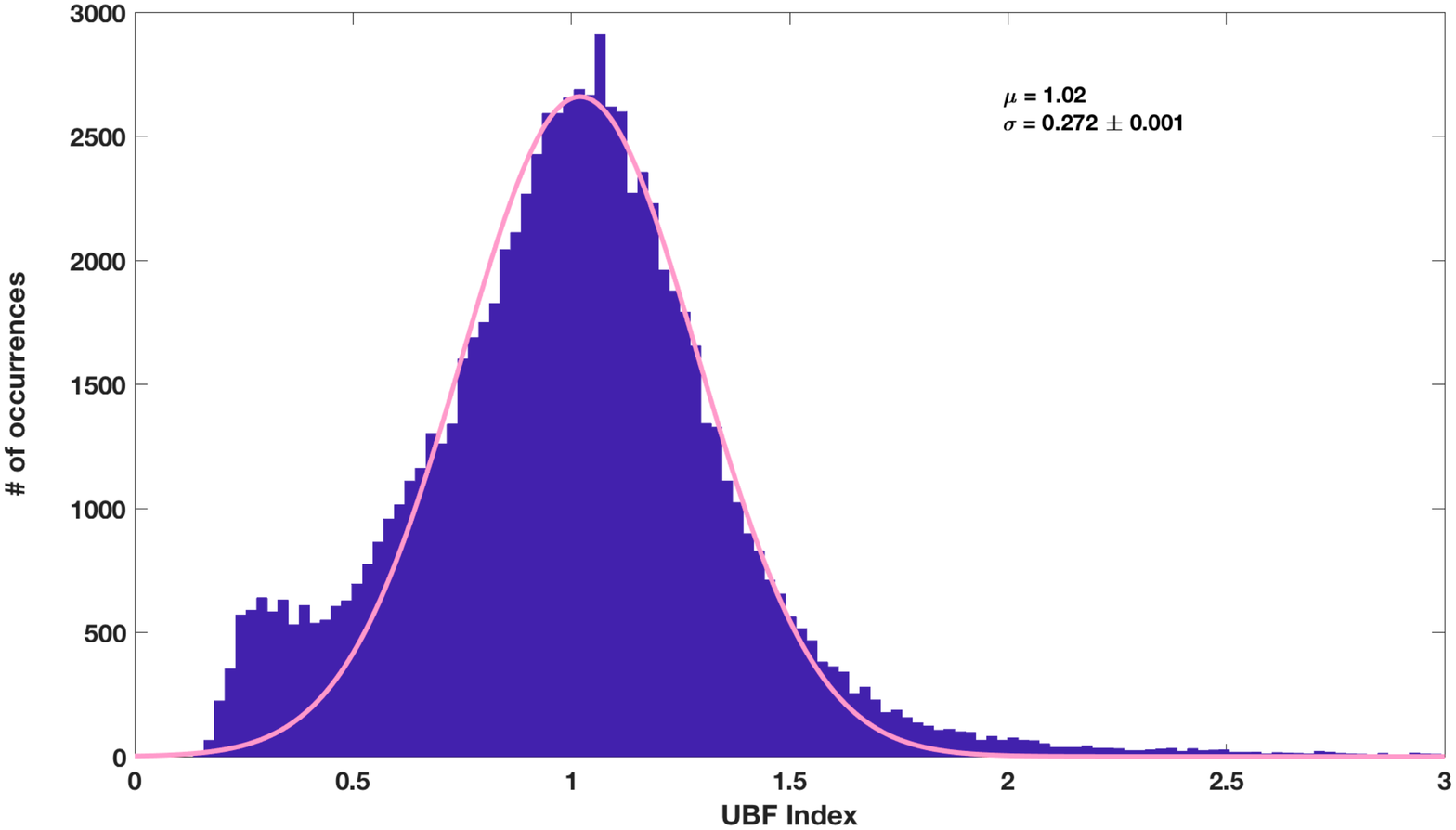}
                       \caption{ Histogram of occurrences of the UBF index. 
                       The fit to a normal distribution is shown by a pink line; dispersion is indicated in the inset.}
                       \label{UBFI_dist}
                     \end{figure}     
                     
The UBFI histogram of occurrences is shown in Figure 7. It displays a maximum at UBFI~$ \simeq 1$ 
that can be well fitted to a normal distribution with dispersion $\sigma = 0.272 \pm 0.001$ that shows the
goodness of   fit  in Eq. 3.  Some excess is observed
in the wings of the distribution produced by the outliers, the weak nebular structures we want to detect 
through the UBF index.  We note that this trend indicates that the galactic UV background fluctuations are not fully
subtracted out in the GALEX image processing pipeline.

There is also a secondary maximum at UBFI $< 0.5$ that comes from
the areas with BSVR $\leq 0.6$ in Figures 5 and 6. In these areas
\begin{equation}
\frac {\sum _{l=1}^{l=10} (Maximum(l)-Mean(l))}{\sum _{l=1}^{l=10} (Mean(l)-Minimum(l))} < 0.6
\label{f1}
\end{equation}
\noindent
or
\begin{equation}
\frac{\left< Maximum \right> - \left< Minimum \right>}{\left< Mean \right>}  < 1.6 
\label{f3}
,\end{equation}
\noindent
{i.e.},  these index values trace areas of the  sky where the UV background is unusually uniform. The pattern is most clearly recognizable in equatorial coordinates,
both the Earth's equator and some polar orbits  are readily recognizable (see Figure 8). 

The main conclusion to be drawn is that   {the UV background is characterized by 
a natural level of fluctuation that is artificially damped in certain areas of the sky}. This damping   most 
likely results from the GALEX data processing.

                     \begin{figure*}[h]
                     \centering
                       \includegraphics[width=12cm,angle=270]{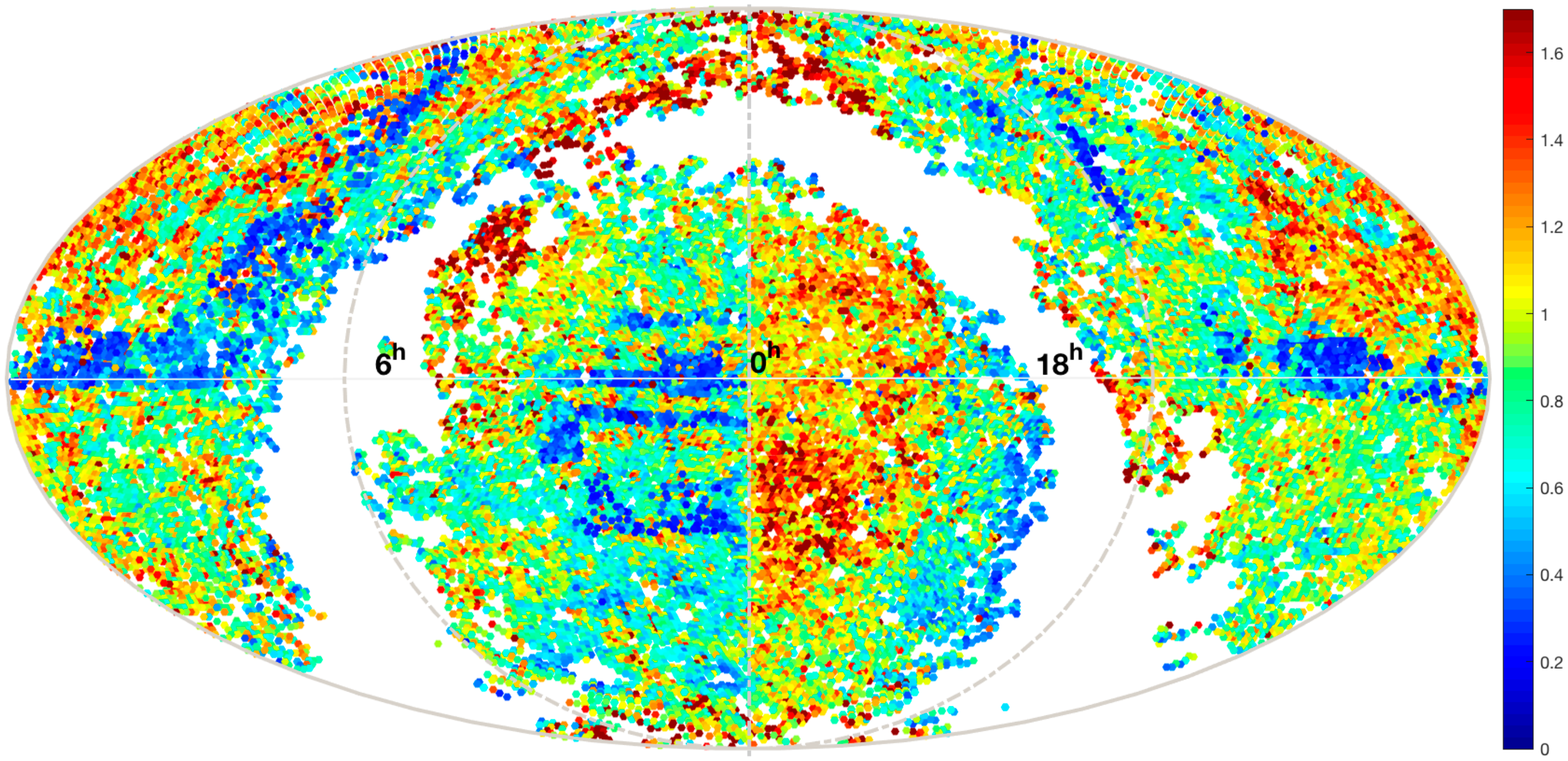} 
                        \caption{ All-sky map of the UBF index in equatorial coordinates (ICRS) and  Aitoff projection. 
                        Shown is the band with low UBFI at the equator.  }
                       \label{UBFI_galaxy}
                       
                     \end{figure*}

                     \begin{figure*}[h]
                     \centering
                       \includegraphics[width=8cm,angle=270]{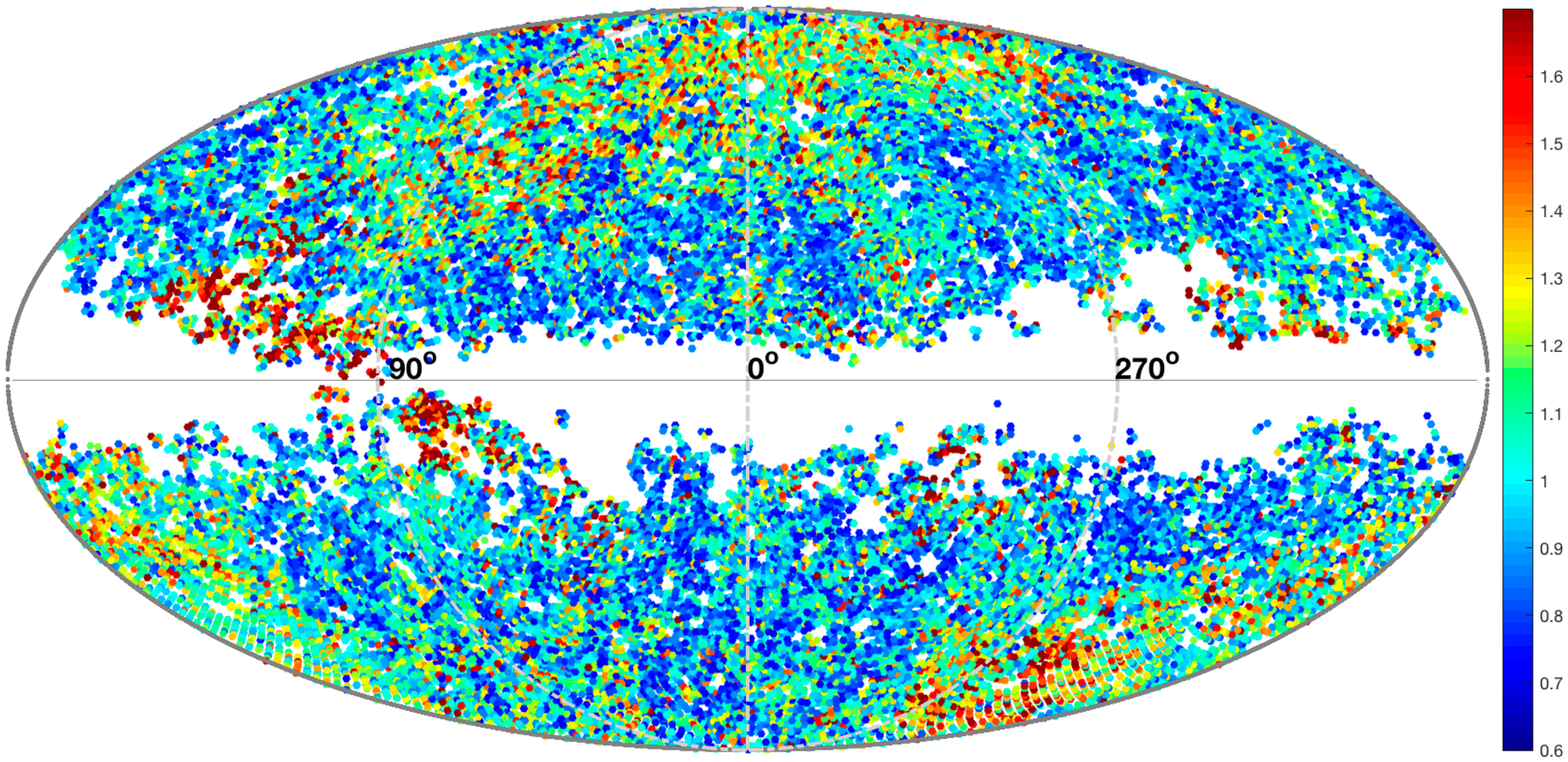} 
                       \includegraphics[width=8cm,angle=270]{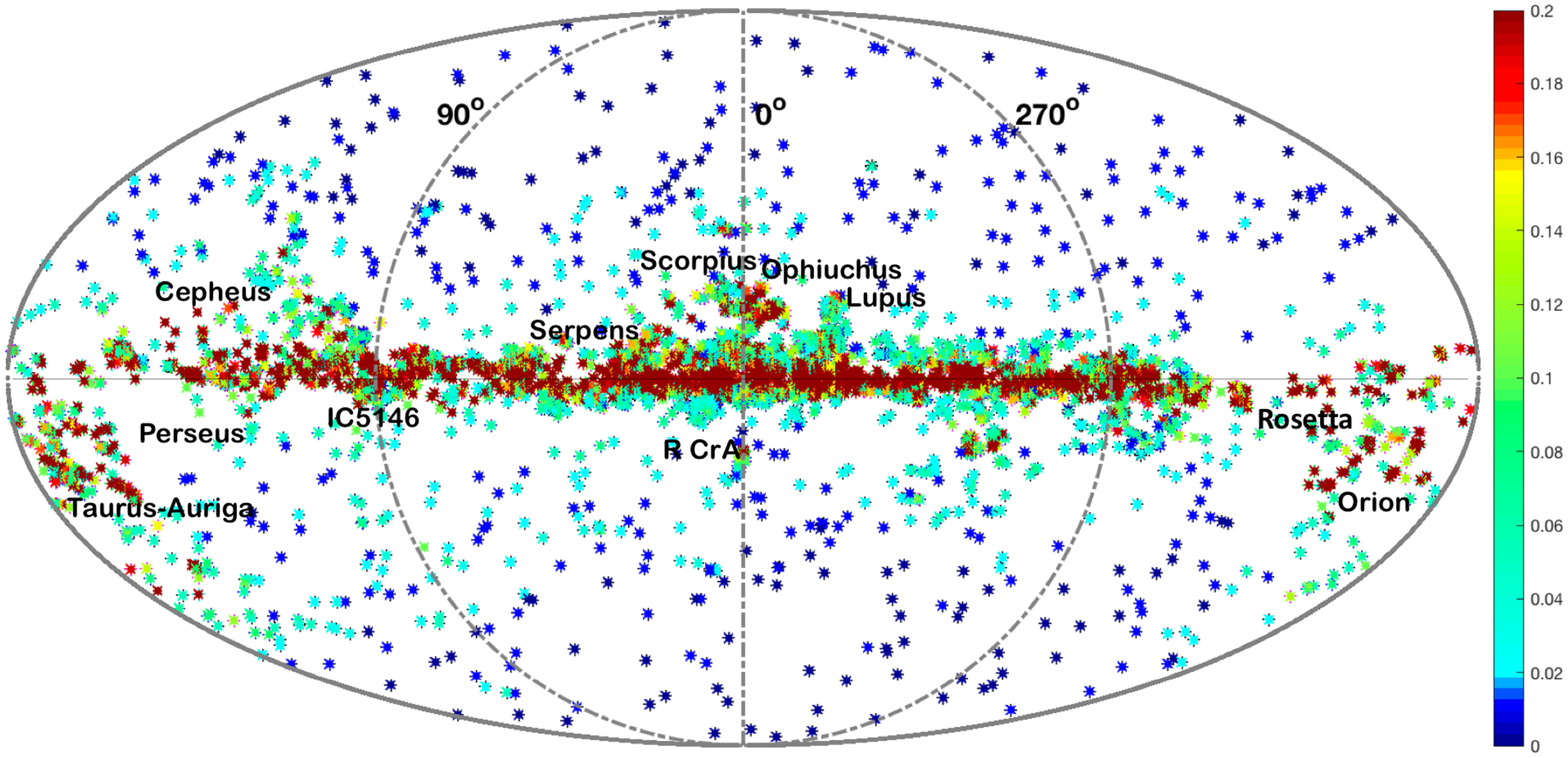} 
                       \includegraphics[width=8cm,angle=270]{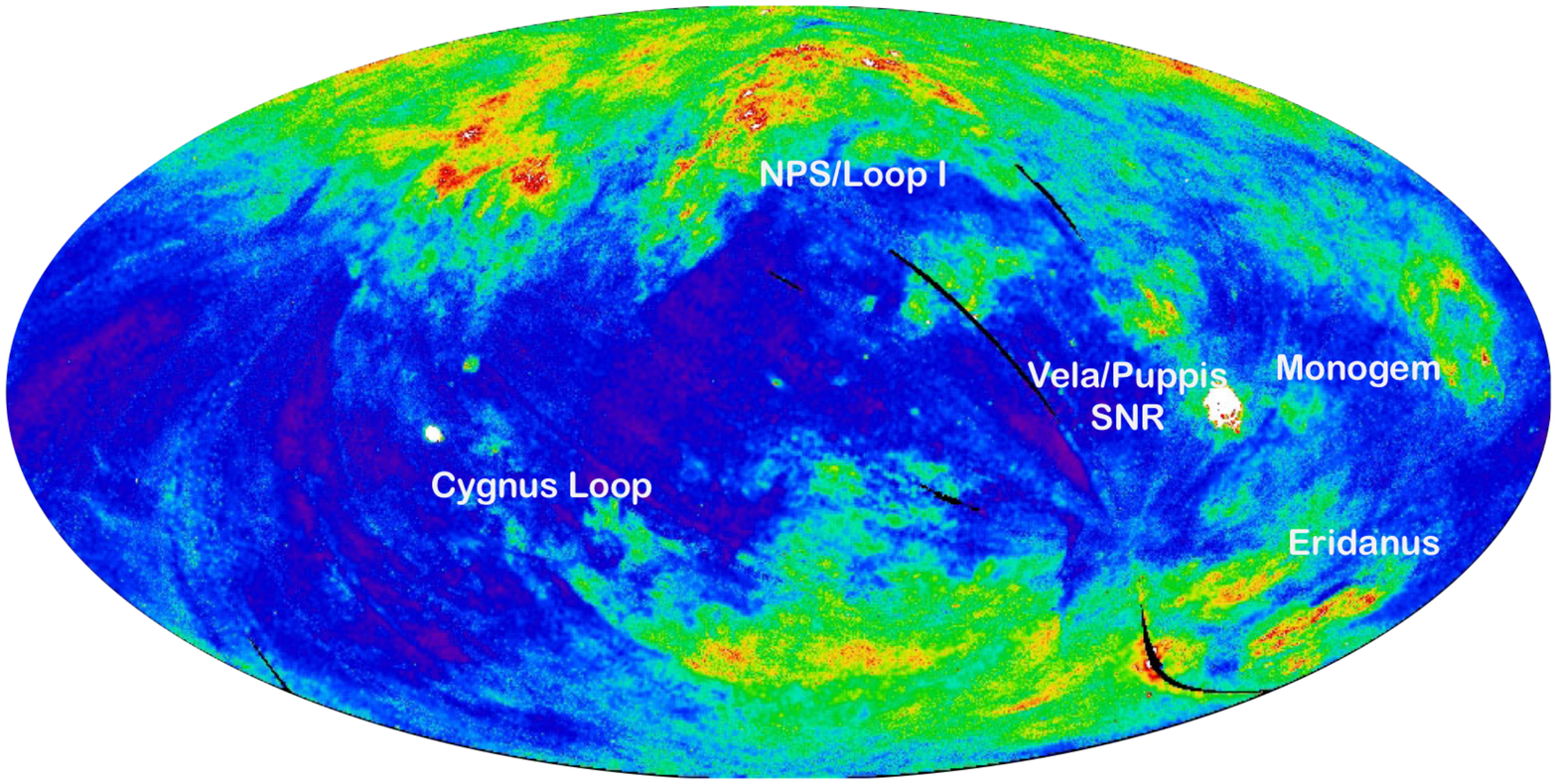}
                        \caption{ {\bf Top:} Map of the UBF index in the area surveyed by GALEX. {\bf Middle:} Location of the 
                        galactic dust clouds according to DB2002; the E(B-V) values from the catalog are color-coded. {\bf Bottom:}
                        Soft X-ray background measured by ROSAT (Snowden et al. 1995). }
                       \label{UBFI_galaxy}
                       
                     \end{figure*}

\section{Results: UV background fluctuations derived from the GALEX FUV AIS}

The UBF index all-sky map is built from the 83.081 GALEX AIS FUV tiles and represented in  galactic coordinates (Aitoff projection)  in Figure 9 (top panel). 
The strongest fluctuations are observed around $l=90^o$ towards Cepheus and the IC5146 star forming regions. 
The comparison with the Dutra and Bica catalog (Dutra \& Bica, 2002, hereafter DB2002) of galactic dust clouds shows a high degree of 
agreement. DB2002 results from the compilation of 21  catalogs that include from dense molecular clouds to the diffuse infrared cirrus 
in the Magnani et al. (1985) catalog.  In Figure 10 (middle panel), the location of the clouds is represented and the E(B-V), as per
DB2002 work, is color-coded. Star forming regions like the Taurus-Auriga complex or Orion  are readily recognizable in the UBFI 
map as already expected from previous works (see G\'omez de Castro et al. 2015a,b and Beitia-Antero \& G\'omez de Castro 2017, for more details). 

Additional areas with strong fluctuations are at high galactic latitudes, some seem to be associated with prominent ISM shells or loops 
such as the Eridanus loop or  Loop I that are especially noticeable in the soft X-ray background map of the Galaxy built by the ROSAT mission (Snowden  et al. 1995). 

The catalog with the UBF index computed for the 83.081 GALEX AIS FUV tiles is available online at the Joint Center for Ultraviolet Astronomy (jcuva.ucm.es) and through the services of the Centre de Donne\'es Stellaires (CDS). For each GALEX AIS tile, the galactic and equatorial (ICRS) coordinates of the  center of the field, as well as the UBF index are provided (see Table 1, for an excerpt). 
We note that there typically 2 or 3 entries per location in the sky since the AIS mapped  the whole sky several times. As shown in the table,
in most cases the UBFI is the same for all measurements of the same field. However, it should be noted that we occasionally   detected discrepancies
(see the first two entries in Table 1). These values are left as calculated in the catalog for reference.

\begin{table}[]

  \centering
  \caption{UBF index from GALEX/AIS FUV images.}
  \small
  \label{tablaresumen}      
    \begin{tabular}{ccccc}
    \hline
     $l_{gal}$ & $b_{gal}$ & R.A. & Declination & UBF index \\  
     ($deg$) & ($deg$) & ($deg$) & ($deg$) & \\
      \hline \hline
0.022187519 &   -70.97773424    & 32.641        & -36.056       & 0.931 \\
0.022187519 &   -70.97773424    & 32.641        & -36.056       & 1.061 \\
0.032145913 &   -50.73732407    & 57.803        & -41.204       & 0.864 \\
0.032145913 &   -50.73732407    & 57.803        & -41.204       & 0.864 \\
0.042265762 &   22.17150569     & 138.84        & -15.966       & 1.018 \\
0.042265762 &   22.17150569     & 138.84        & -15.966       & 1.018 \\
0.045154008 &   8.719991456     & 127.41        & -24.086       & 0.373 \\
0.045154008 &   8.719991456     & 127.41        & -24.086       & 0.373 \\
0.061980053 &   -66.12334178    & 38.328        & -37.758       & 1.212 \\
\hline
\end{tabular}
\end{table}

\section{Interstellar clouds in the far UV}

\begin{figure}[h]
\centering
\includegraphics[width=4cm]{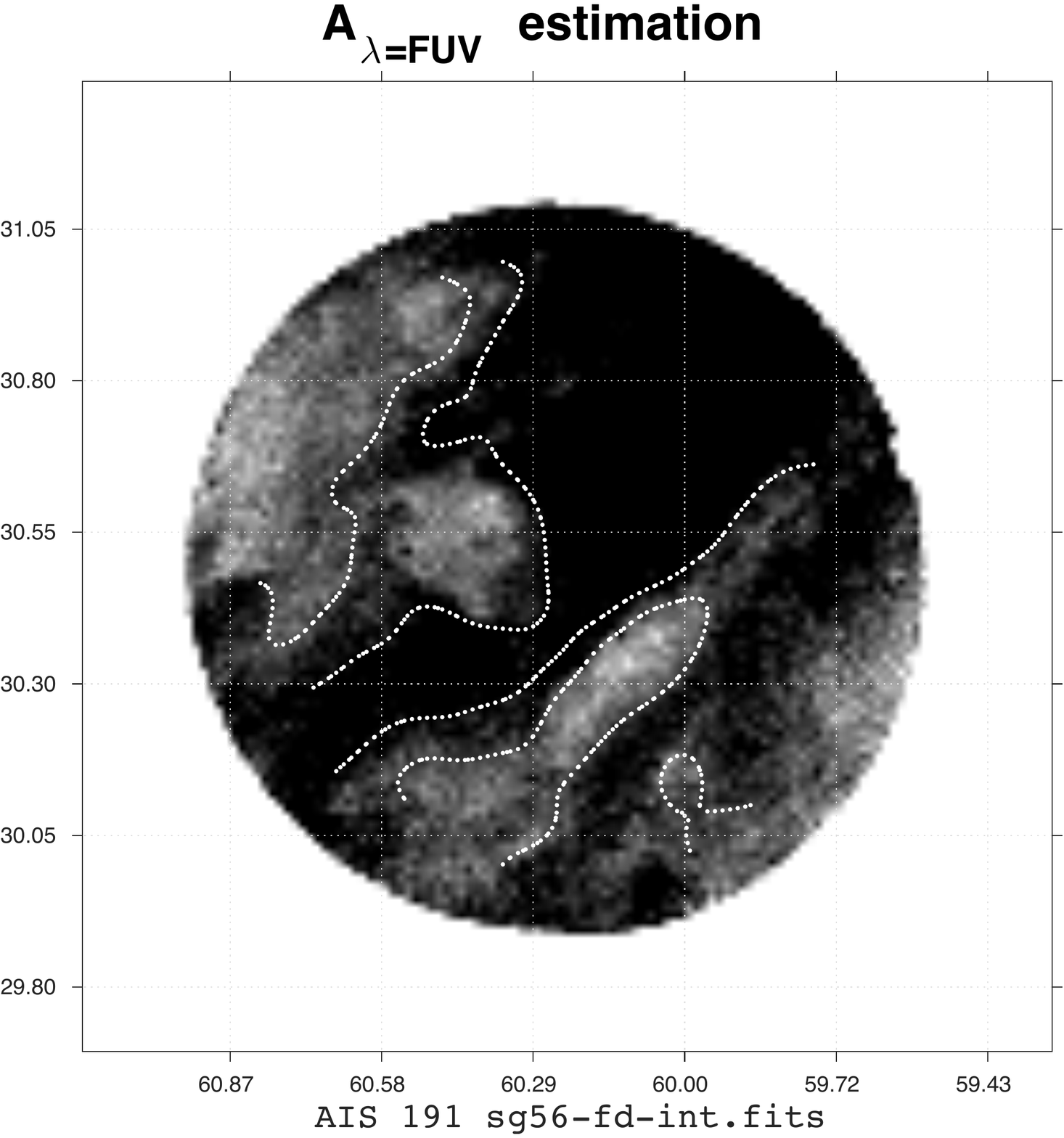}\hspace{0.2cm}\includegraphics[width=4cm]{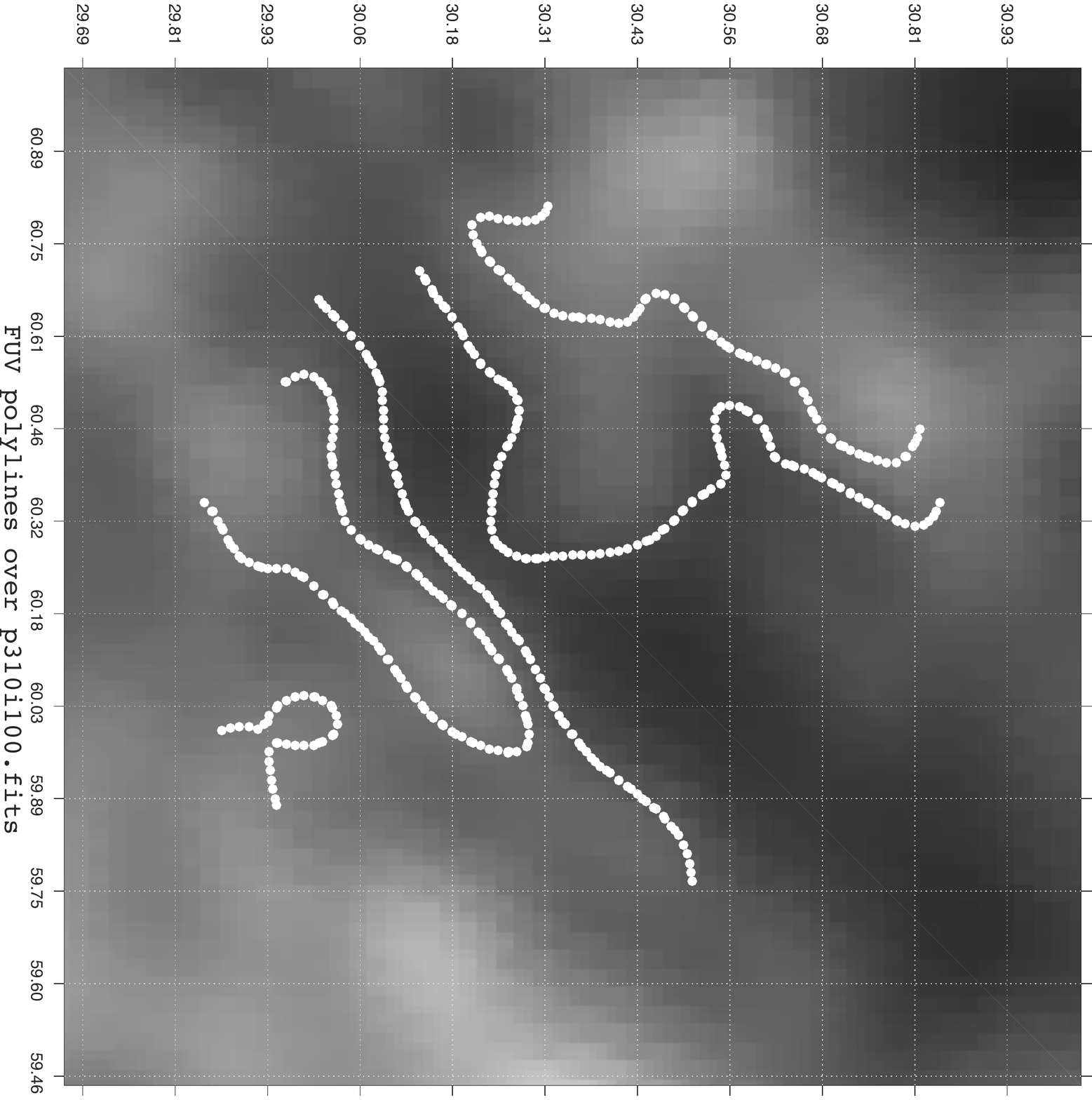}
\includegraphics[width=4cm]{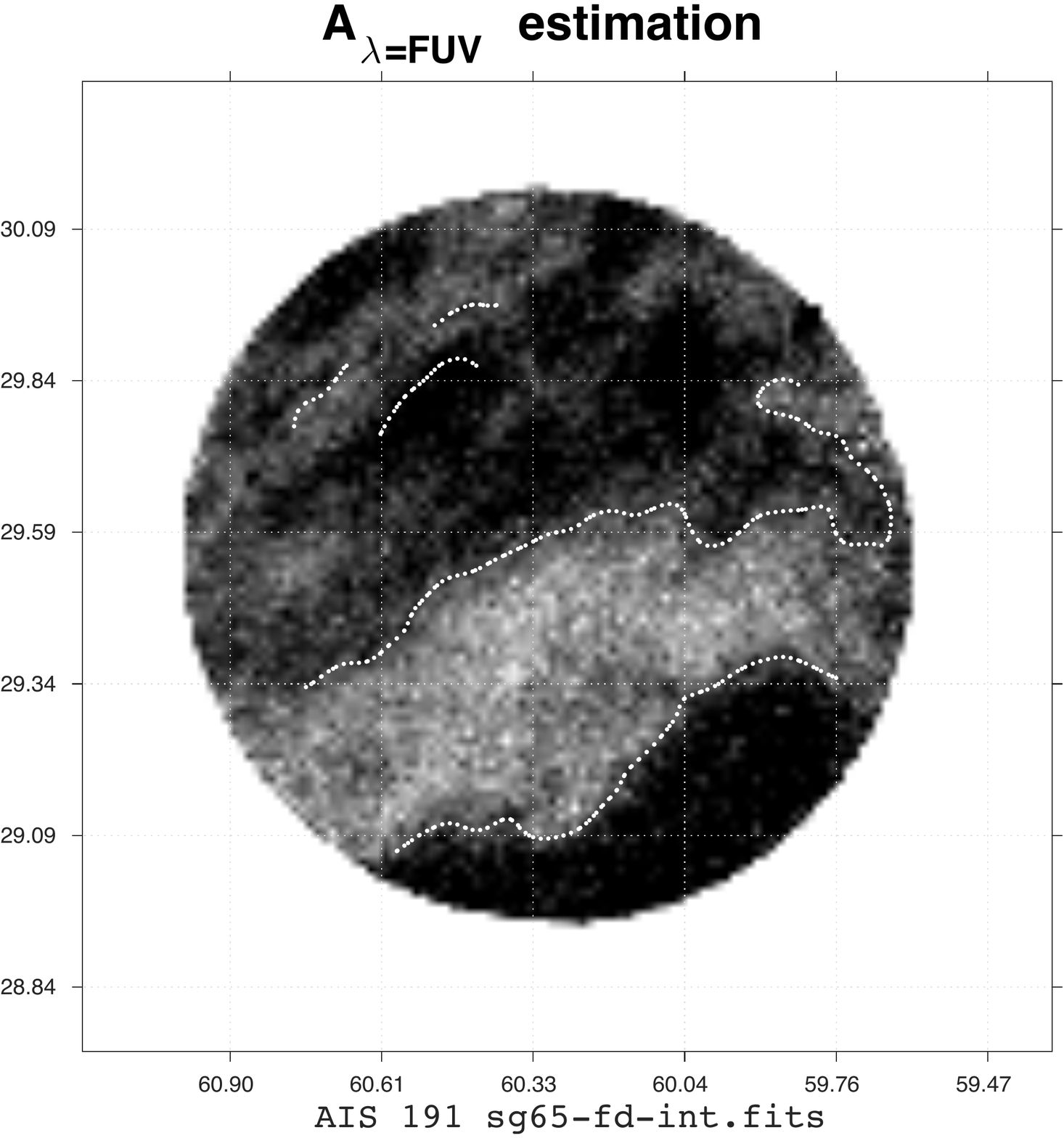}\hspace{0.2cm}\includegraphics[width=4cm]{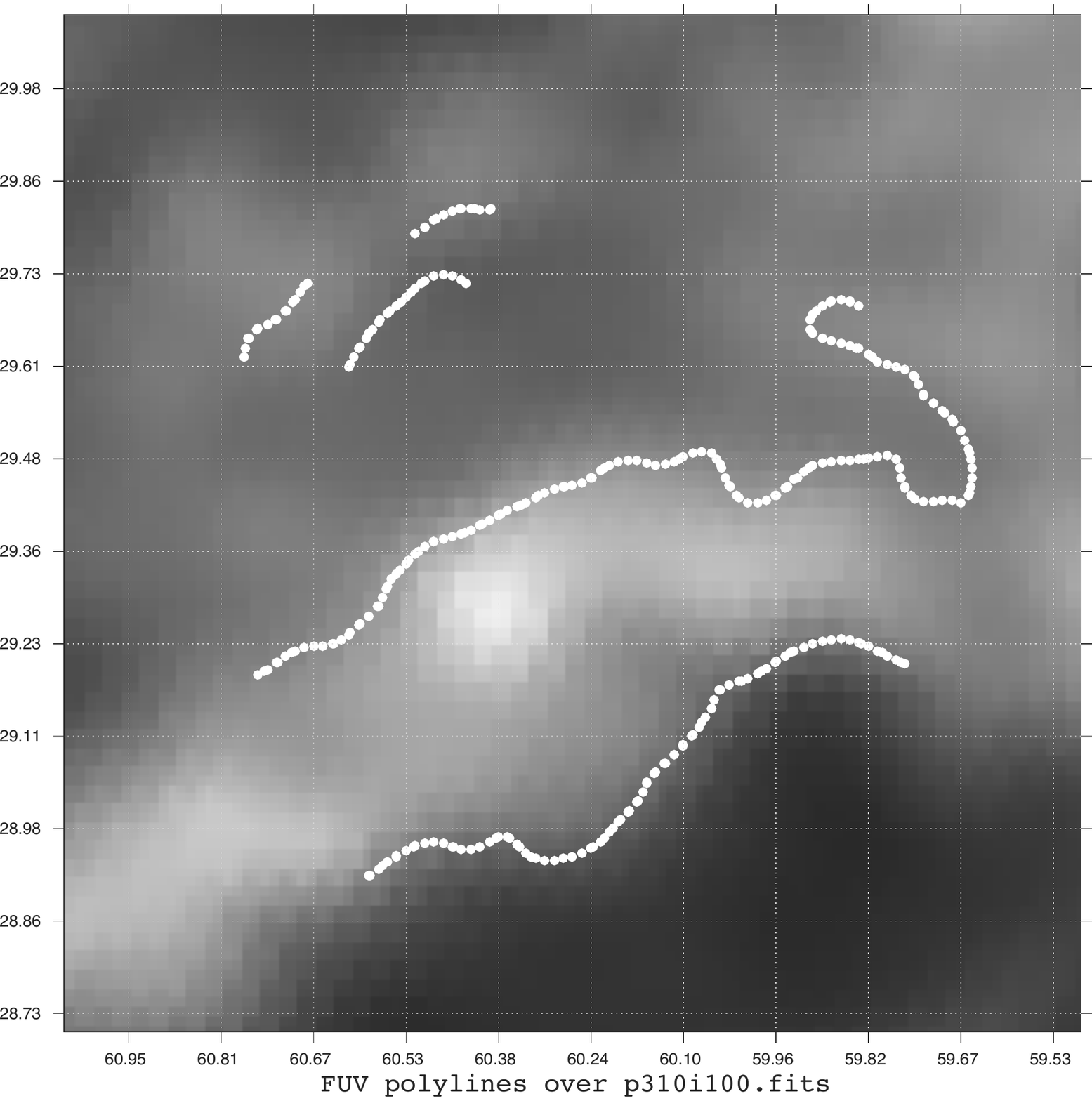}
\includegraphics[width=4cm]{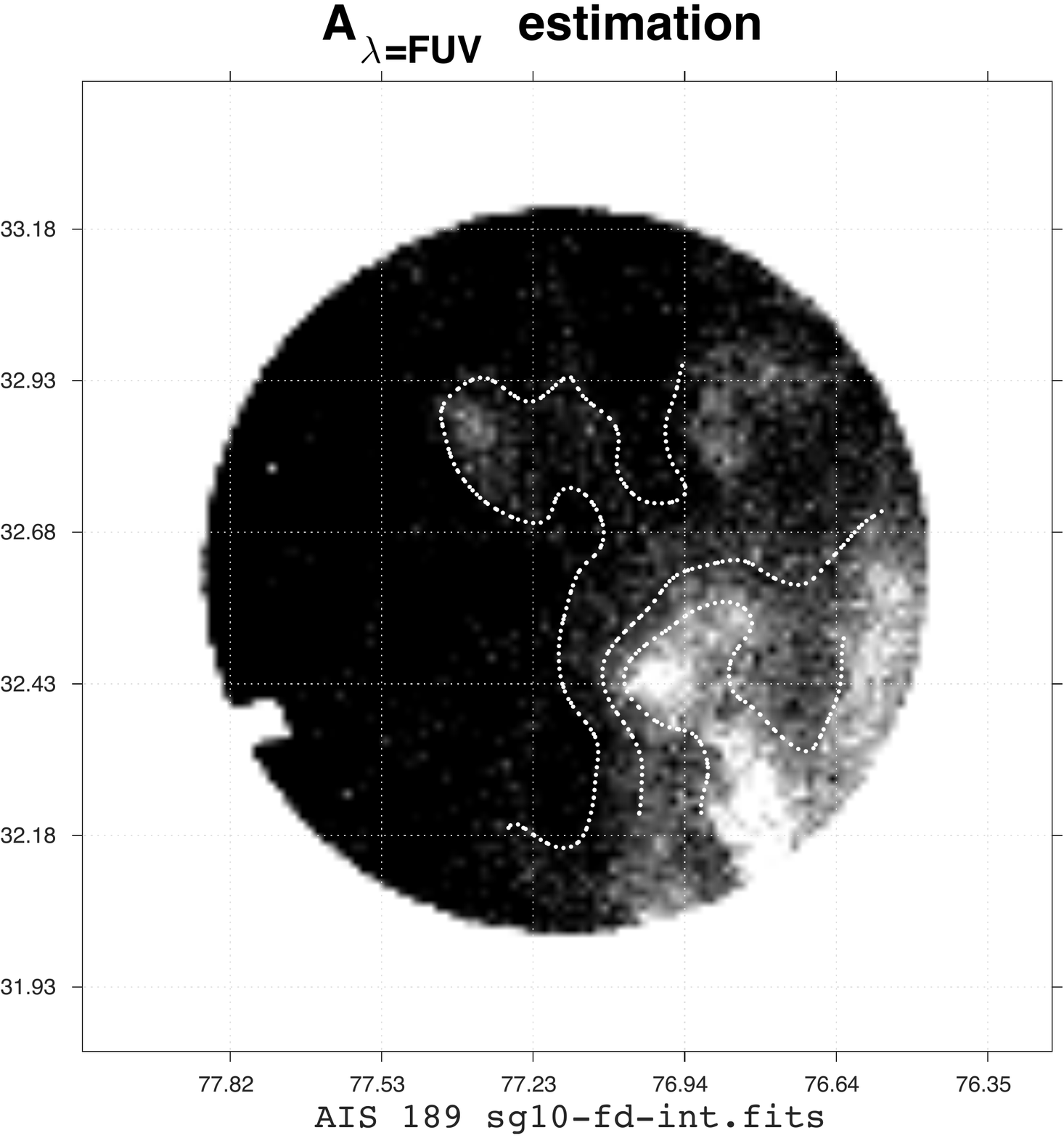}\hspace{0.2cm}\includegraphics[width=4cm]{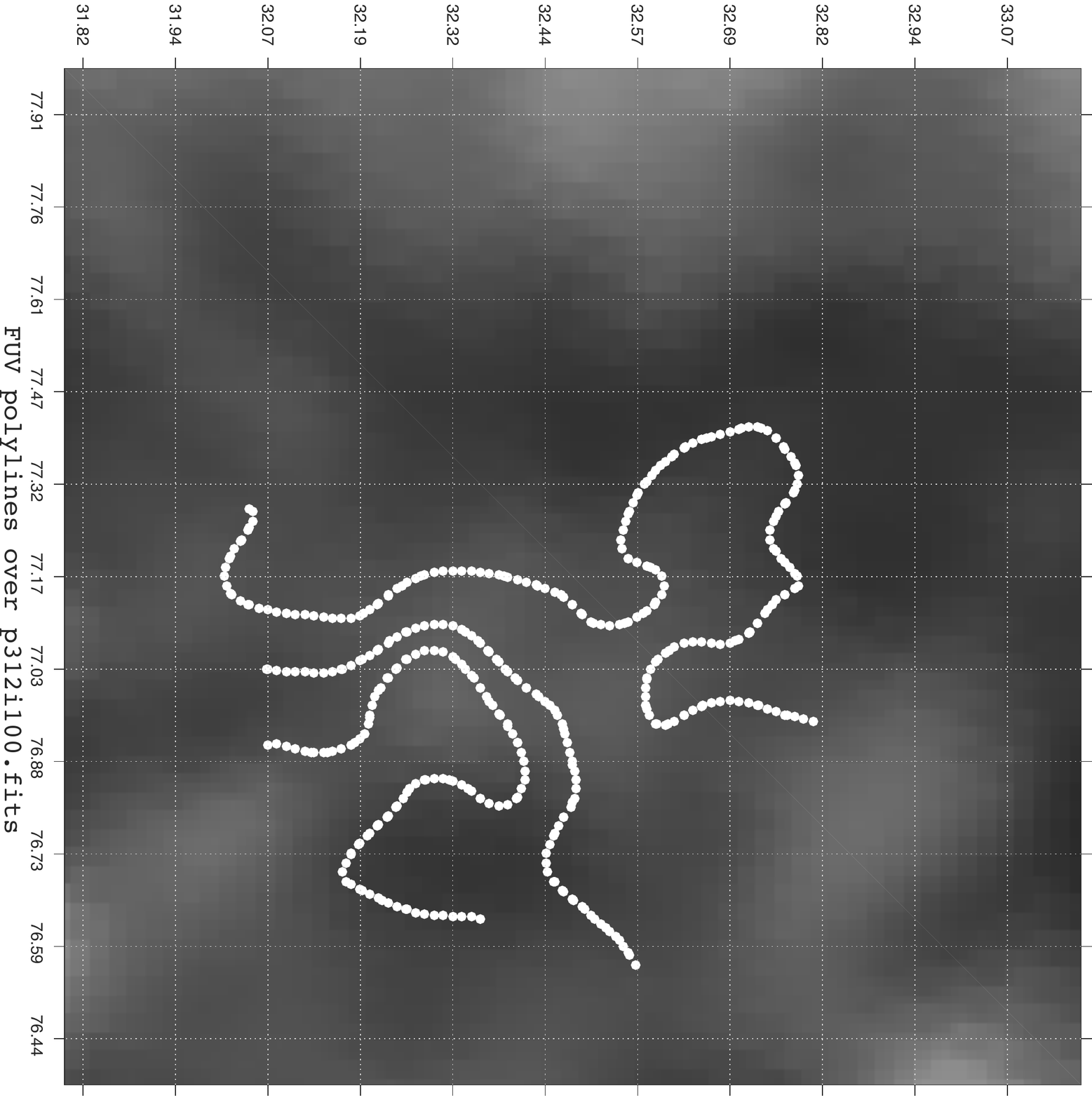}
\includegraphics[width=4cm]{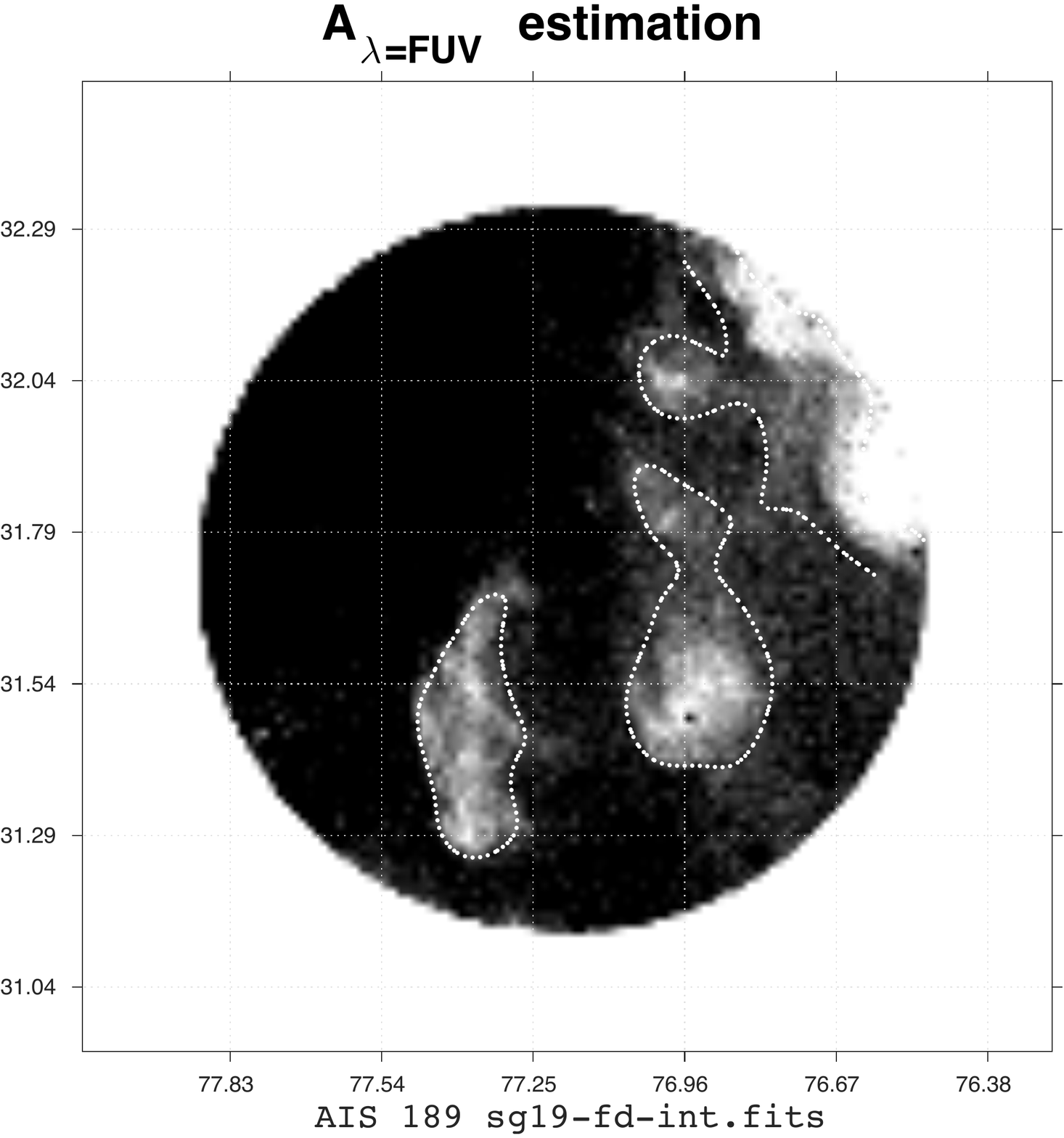}\hspace{0.2cm}\includegraphics[width=4cm]{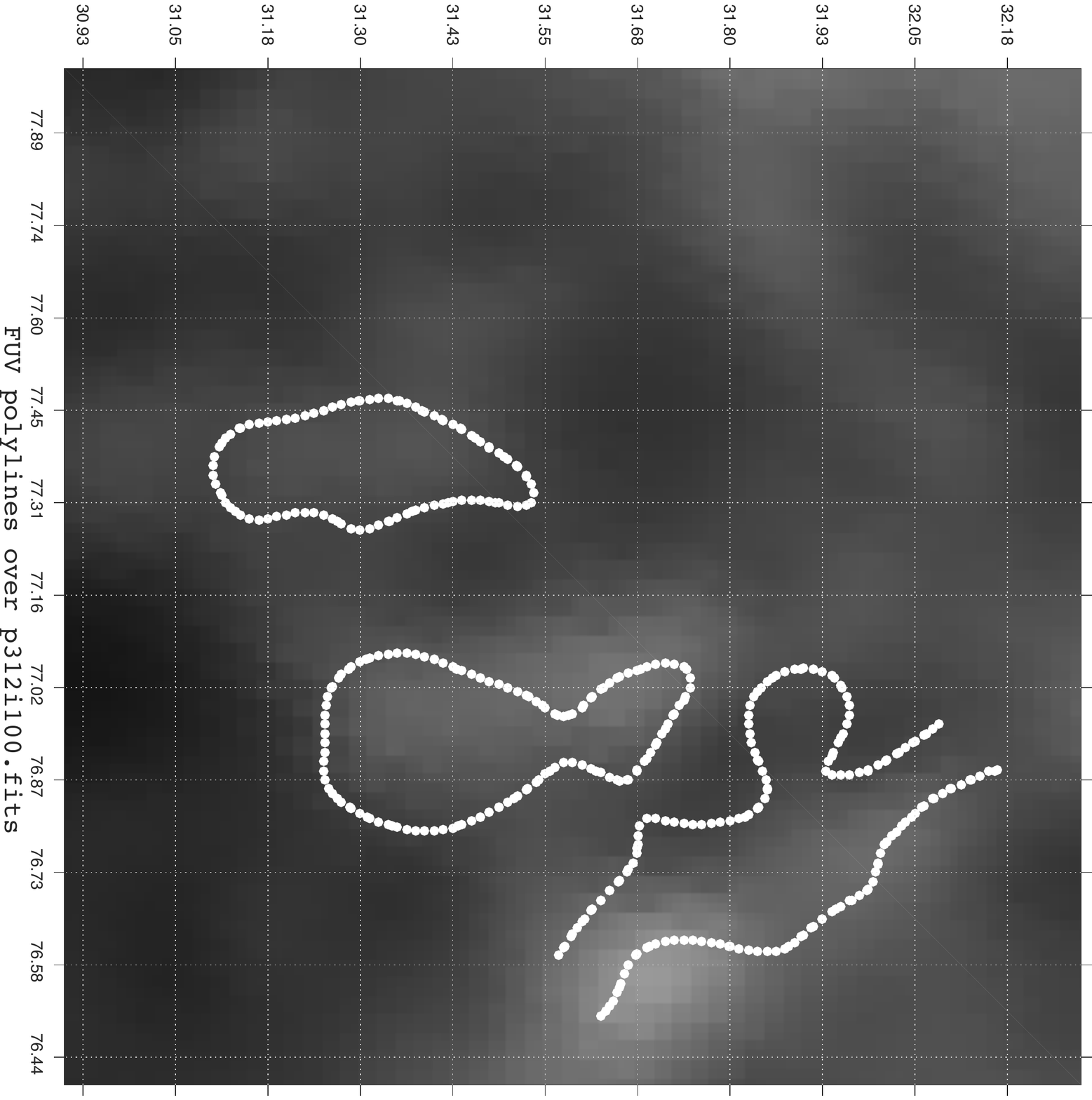}
\caption{{\bf Left:} Relative extinction maps for some areas of the Taurus-Auriga star forming complex. {\bf Right:} The same areas in the 
  IRAS 100 microns maps. The main features extracted from the UV relative extinction maps are marked in both plots to test the overlap
  with IRAS; in these examples the match is significant.}
\label{fig:polylines}
\end{figure}

According to the histogram in Figure~7, the fluctuations introduced by interstellar structures 
are expected to be at UBFI~$>1.3$, which corresponds to $2.5 \sigma$ above the average (see also Figure 9a).
In general, good matches are found between the  FUV structures and the infrared structures in the nearby molecular clouds.
A good example is the Taurus-Auriga region. In figure 7, the FUV absorption is compared with the infrared
emission for some dark globules and filaments in the area. In the left panels, the FUV maps are displayed 
in logarithmic scale; they are relative-extinction maps computed  as
        \begin{equation}
          A _{i,j} = - 2.5 log_{(10)} (C_{i,j} / \langle C^{0}_k \rangle )
          \label{draine_eq}
        ,\end{equation}
where $C_{i,j} $ is the count rate in resel $(i,j),$ and $ \langle C ^0 _k \rangle $ is the average count rate in the tile. 
In the right panel, IRAS images (100 microns)  of the same features are plotted. Key contours from the left panel (FUV) 
are overimposed in the right panel for guidance. We note that the FUV extinction
$A_{FUV}$  scales with the column density of dust grains\footnote{For a given extinction law, $A_{FUV}/A_V$ is constant
and $N_H/A_V \approx 1.8  \times 10^{21}$~atoms cm$^{-2}$ mag$^{-1}$ (Bohlin et al. 1978) for the ISM.} $N_d$  as A$_{\rm FUV} \propto N_d$, and the thermal emission $F_{100 \mu}$ is also expected to be proportional to  $N_d $, thus roughly  $A_{FUV} \propto F_{100 \mu}$. In practice, the details of the grain composition, albedo, and size distribution will introduce deviations over this simple trend providing
a unique tool for the characterization  of  dust grains.  

\begin{figure}[h]
\centering
\begin{tabular}{cc}
\includegraphics[width=6cm]{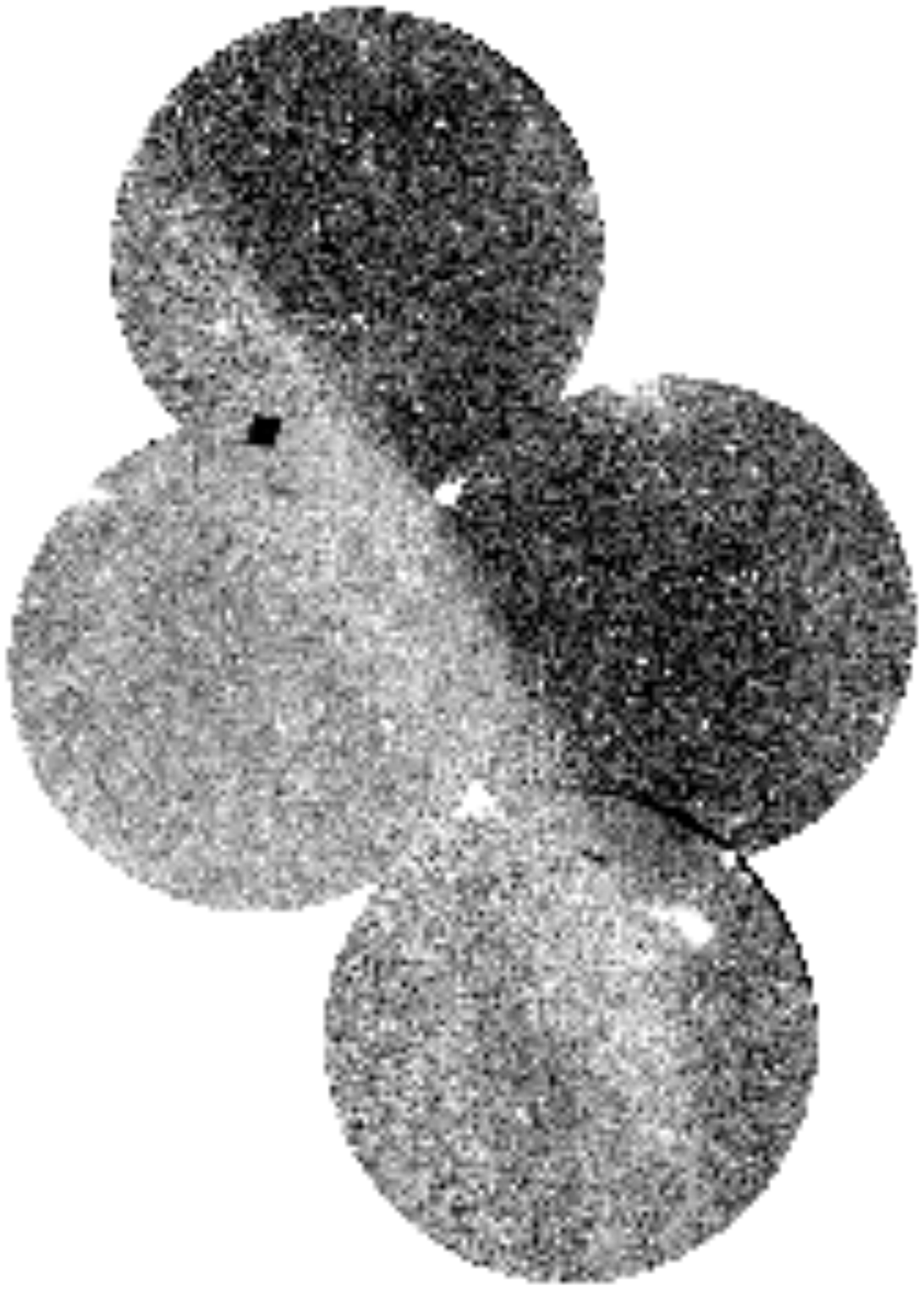} & \includegraphics[width=3cm]{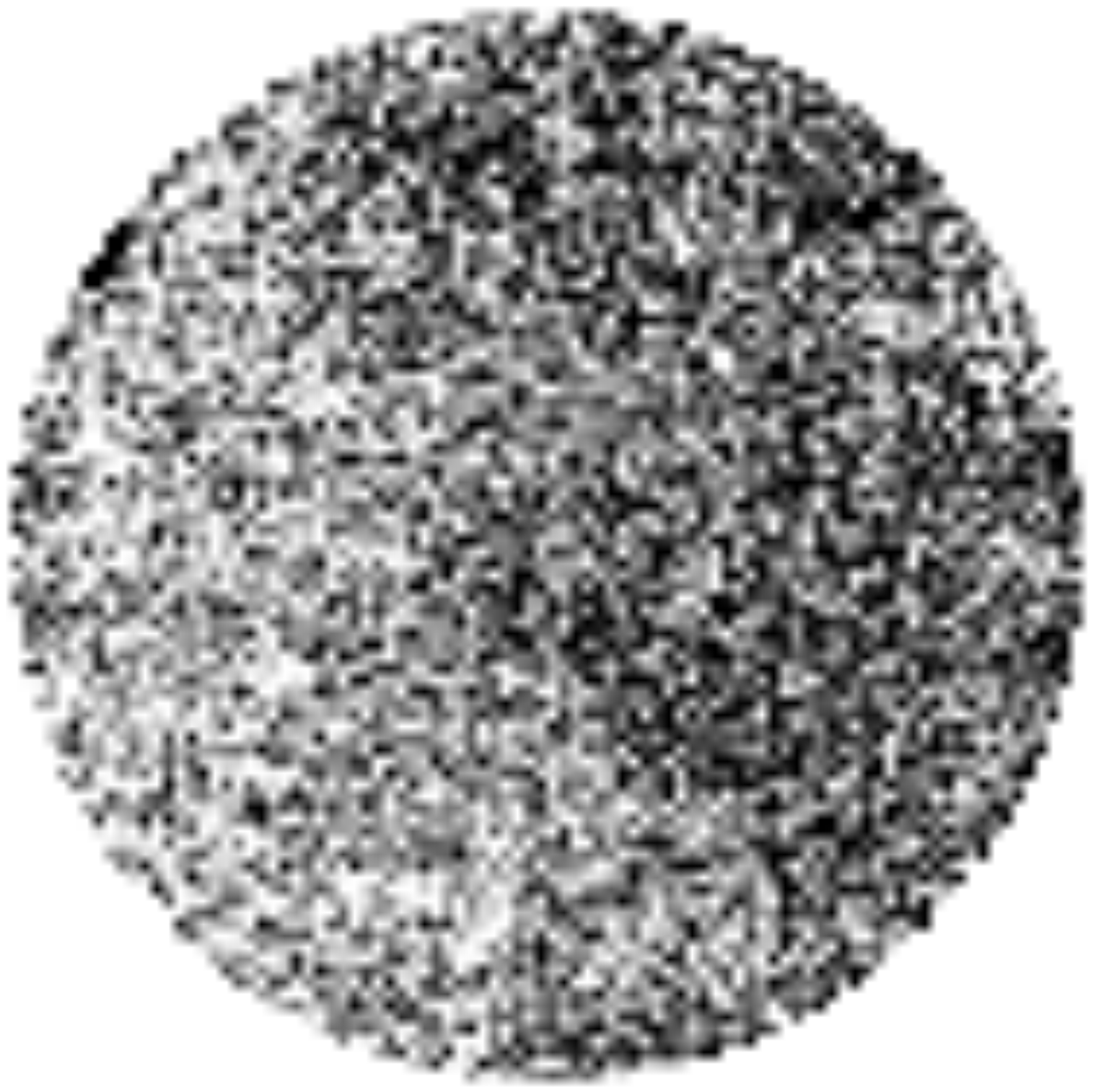}\\
\includegraphics[width=6cm]{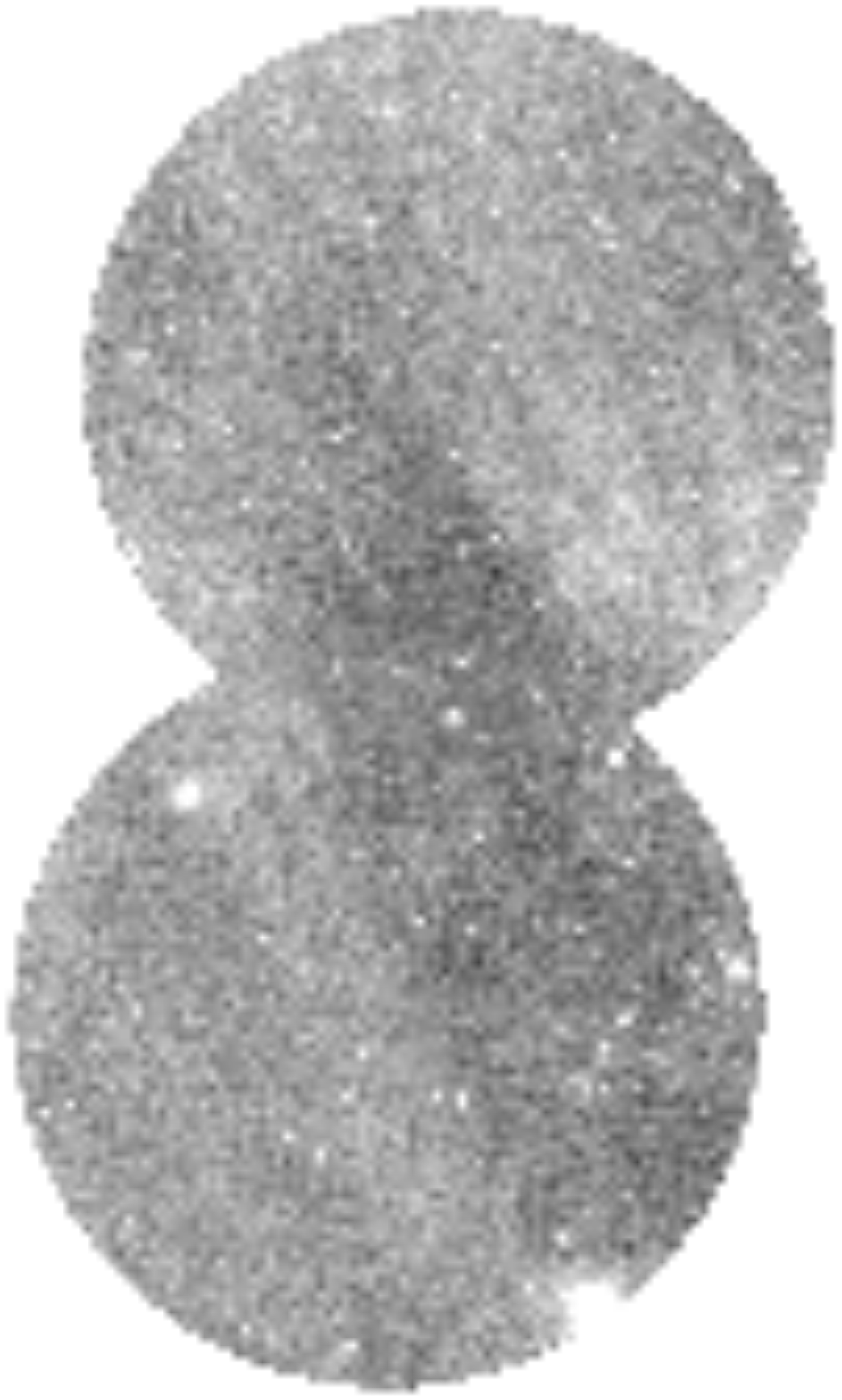} & \includegraphics[width=3cm]{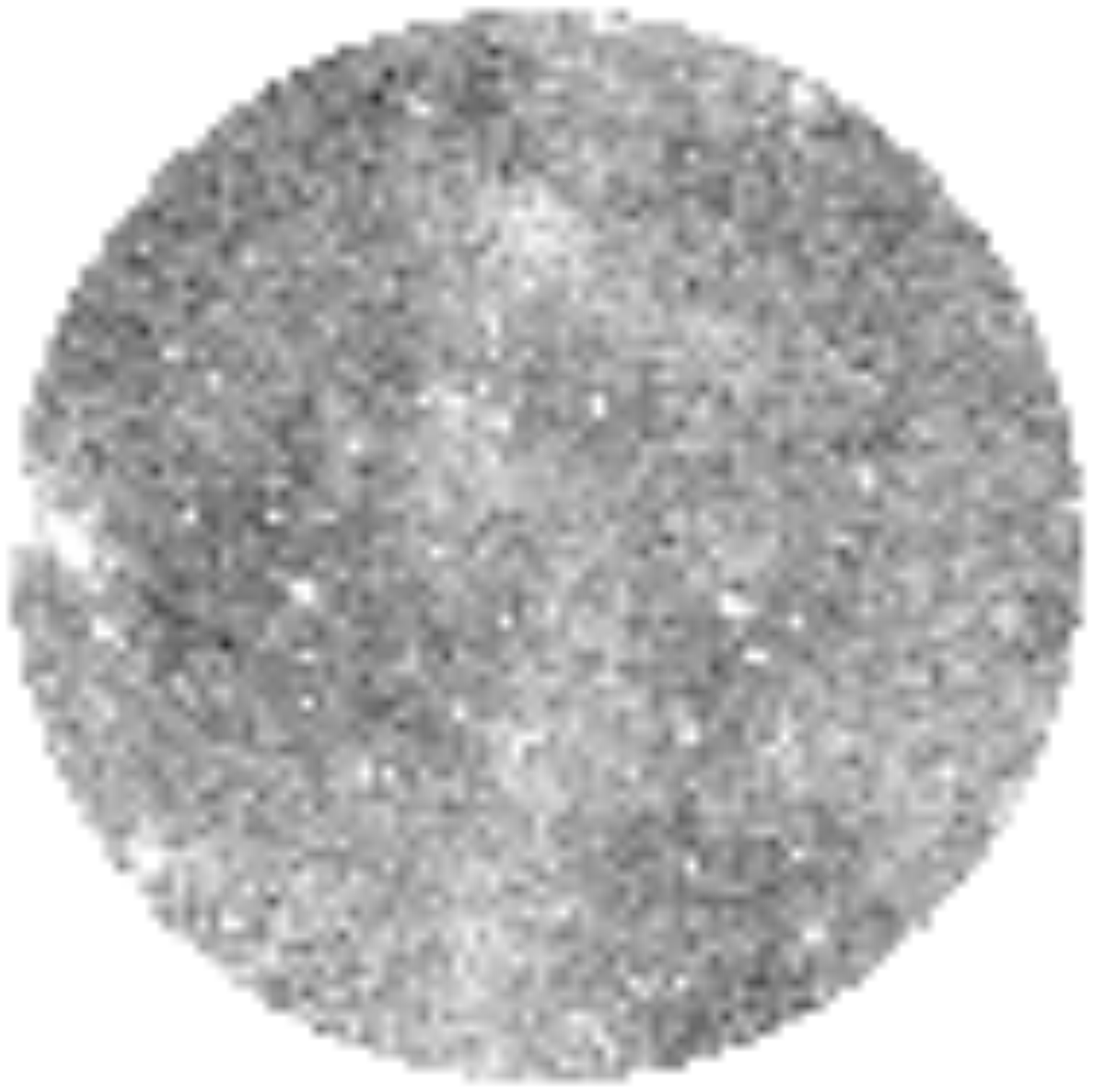}\\
\end{tabular}
 
\caption{Four of the best matches with H~I clouds from the  GALFA-H~I survey. According to  Begun et al. 2010, 
(a), (b), (c), and (d) correspond to entries 5, 15, 6, and 54 in  Table 1 of  the catalog.}

\label{4XneutralHI}
\end{figure}

There is not, however, a similar good match with the diffuse structures detected by the Galactic Arecibo L-Band Feed Array H~I (GALFA H~I) survey. GALFA~HI maps with unprecedented sensitivity and spatial resolution of 3.5~arc~min, the distribution of neutral hydrogen in the 21 cm line (Begun et al. 2010). Ninety-six H~I clouds were reported to have LSR velocities smaller than 90 km s$^{-1}$ and, in principle,  could 
be close enough to be detectable within the GALEX AIS. Although some   good matches have been found, as shown
in Figure 11, most of the clouds have not been detected by GALEX, probably because of the limited depth of the GALEX AIS survey.

\section{Conclusions}
\label{ending}

In this work, we  mined the FUV images in the GALEX all-sky survey to detect weak extended structures in the ISM.
We  derived an index, the UBF index, that can be reliably used to search for fluctuations of the FUV background in the sky. 
The UBFI is greater than  1.3 in the main nearby star forming complexes and some prominent loops in the ISM.

\begin{acknowledgements}
This work has been partly support by grants ESP2015-68908-R and ESP2017-87813-R from the Ministry of Science, Innovation and
Universities of Spain.
\end{acknowledgements}



\begin{thebibliography} {}

\bibitem{aa} Armengot, M., Sanchez, N., Lopez-Santiago, J. et al.,  2014, ApSS, 354, 113

\bibitem{bb} Begum, A., Stanimirovi{\'c}, S., Peek, J.~E. et al. 2010, ApJ, 722, 395 

\bibitem{bc} Beichman, C. A., Neugebauer, G., Habing, H. J.,  Clegg, P. E., Chester, T. J., eds. (1988). Infrared Astronomical Satellite (IRAS): Catalogs and Atlases (PDF). Volume 1: Explanatory Supplement (2nd ed.). NASA Scientific and Technical Information Division.

\bibitem{bd} Beitia-Antero, L. and Gomez de Castro, A.~I., 2017, MNRAS, 469, 2531 

\bibitem{be} Bianchi, L., 2014, ApSS, 354, 103

\bibitem{dd} Draine, B.T., 2003, ApJ, 598, 1017

\bibitem{du} Dutra, C.~M. and Bica, E., 2002, A\&A, 383, 631

\bibitem{gg} Gomez de Castro, A.~I., Lopez-Santiago, J., Lopez-Martinez, F. et al., 2015a, ApJS, 216, 26 

\bibitem{gu} Gomez de Castro, A.~I., Lopez-Santiago, J., Lopez-Martinez, F. et al., 2015b, MNRAS, 449, 3867G 

\bibitem{ha} Hamden, E.~T., Schiminovich, D. and Seibert, M., 2013, ApJ, 779 180 

\bibitem{ma} Magnani, L., Blitz, L. and Mundy, L., 1985, ApJ, 295, 402

\bibitem{mm} Martin, D.~C., Fanson, J., Schiminovich, D. et al., 2005, ApJL, 619, L1 

\bibitem{mo} Morrissey, P., Conrow, T., Barlow, T.~A. et al., 2007, ApJS, 173, 682

\bibitem{mu1} Murthy, J., Henry, R.~C. and Sujatha, N.~V., 2010, ApJ, 724, 1389
 
\bibitem{mu2} Murthy, J., 2014, ApJS, 213, 32

\bibitem{sa} Saslaw, W. C.,  Gaustad, J. E., 1969, Nature, 221, 160S

\bibitem{sn} Snowden, S. L., Freyberg, M. J., Plucinsky, P. P. et al. 1995, ApJ, 454, 643 

\end{thebibliography}

\end{document}